\documentclass[]{aa}  
\usepackage{graphicx}
\usepackage{natbib}
\usepackage{longtable}
\usepackage[english]{babel}
\usepackage{txfonts}
\usepackage{color}

\begin{document}
   \title{Single-epoch VLBI imaging study\\ of bright active galactic nuclei at 2~GHz and 8~GHz}
   \titlerunning{Single-epoch VLBI imaging of active galactic nuclei at 2~GHz and 8~GHz}

   \author{A. B. Pushkarev
          \inst{1,2,3},
          Y. Y. Kovalev\inst{4,1}
          }

   \institute{Max-Planck-Institut f\"ur Radioastronomie, 
              Auf dem H\"ugel 69, 53121 Bonn, Germany\\
              \email{apushkar@mpifr.de}
         \and 
             Pulkovo Astronomical Observatory, Pulkovskoe Chaussee 65/1, 196140 St. Petersburg, Russia
	 \and
             Crimean Astrophysical Observatory, 98409 Nauchny, Ukraine
         \and
             Astro Space Center of Lebedev Physical Institute, Profsoyuznaya 84/32, 117997 Moscow, Russia\\
             \email{yyk@asc.rssi.ru}
             }

   \date{Received 4 April 2012; accepted 22 May 2012}

 
  \abstract
    {}
   {We investigate statistical and individual astrophysical properties of active galactic 
    nuclei (AGNs), such as parsec-scale flux density, core dominance, angular and linear 
    sizes, maximum observed brightness temperatures of VLBI core components, spectral index 
    distributions for core and jet components, and evolution of brightness temperature along 
    the jets. Furthermore, we statistically compare core flux densities and brightness 
    temperature as well as jet spectral indices of $\gamma$-ray bright and weak sources.}
   {We used 19 very long baseline interferometry (VLBI) observing sessions carried out
   simultaneously at 2.3~GHz and 8.6~GHz with the participation of 10 Very Long Baseline 
   Array (VLBA) stations and up to 10 additional geodetic telescopes. The observations 
   span the period 1998--2003.}
   {We present here single-epoch results from high-resolution radio observations of 370 AGNs. 
   Our VLBI images at 2.3~GHz and 8.6~GHz as well as Gaussian models are presented and analyzed. 
   At least one-fourth of the cores are completely unresolved on the longest baselines of the 
   global VLBI observations. The VLBI core components are partially opaque with the median value 
   of spectral index of $\alpha_{\mathrm{core}}\sim0.3$, while the jet features are usually 
   optically thin $\alpha_{\mathrm{jet}}\sim-0.7$. The spectral index typically decreases along 
   the jet ridge line owing to the spectral aging, with a median value of $-0.05$~mas$^{-1}$. 
   Brightness temperatures are found to be affected by Doppler boosting and reach up to 
   $\sim$$10^{13}$~K with a median of $\sim$$2.5\times10^{11}$~K at both frequencies. The 
   brightness temperature gradients along the jets typically follow a power law 
   $T_\mathrm{b}\propto r^{-2.2}$ at both frequencies. We find that 147 sources (40\%) 
   positionally associated with $\gamma$-ray detections from the $Fermi$ LAT Second Source 
   Catalog have higher core flux densities and brightness temperatures, and are characterized 
   by the less steep radio spectrum of the optically thin jet emission.}
   {}


   \maketitle
%

\section{Introduction}

The long-term VLBI project RDV (Research \& Development -- VLBA) is designed to perform 
observations of bright, flat-spectrum, compact extragalactic radio sources. The project 
started in 1997 under the coordination of NASA and NRAO. The simultaneous observations at 
2.3~GHz and 8.6~GHz are carried out bi-monthly, making up five to six sessions per year 
with the participation of all ten 25~m VLBA antennas and up to ten geodetic stations 
\citep[for detail see][]{Petrov09}. The participation of the southern antennas such as 
HartRAO (South Africa) and TIGO (Chile) allowed us to improve the {\it uv}-coverage of 
low-declination sources. One of the initial goals of the project was the estimation of 
the precise absolute positions of compact sources and the improvement of the International 
Celestial Reference Frame \citep[ICRF;][]{Ma98}, which was extended by the inclusion of 
776 sources \citep{icrf-ext2-2004} and later 3414 radio astronomical objects constituting 
the ICRF2 \citep{ICRF2}. The second goal was to perform to precise geodesy and the 
determination of antenna reference point changes associated with the motion of tectonic 
platforms \citep{Petrov09}. 

We have successfully used the same data for astrophysical studies of active galactic nuclei 
(AGNs), drawing on data from the NRAO archive\footnote{\tt http://archive.nrao.edu}. 
Since the project was initially focused on performing geodesy and astrometry, the primary 
type of information gathered was in the measured phases of the registered signals, whereas 
for us both the phases and amplitudes are equally important for restoring and analyzing the 
images. We contacted the staff of the non-VLBA stations and received valuable information 
about system equivalent flux densities (SEFD) and system temperatures at different epochs 
of observations. We note that our imaging results should help to complete astrometry, i.e., 
improve the accuracy of the absolute positions of the sources taking into account their 
milliarcsecond structure.

\begin{figure}
 \resizebox{\hsize}{!}{\includegraphics[height=0.5\textwidth,angle=-90]{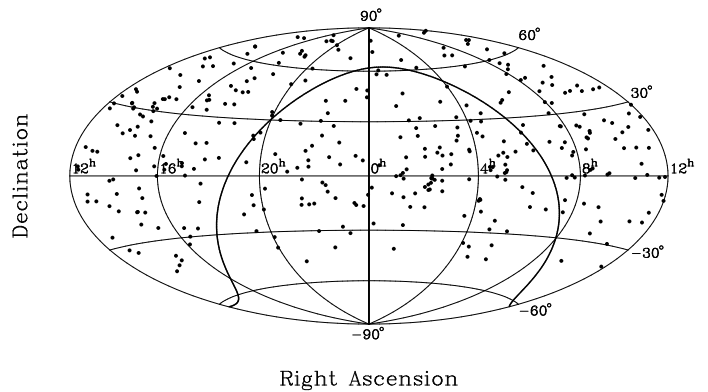}}
 \caption{
          Sky distribution of 370 observed sources in Aitoff equal-area projection of the 
	  celestial sphere in equatorial coordinates. The thick line is the Galactic equator.
	 }
 \label{f:sky}
\end{figure}

In each experiment, about 100 active galactic nuclei are scheduled. The total number of sources 
within RDV project is currently approaching 1000. The main selection criteria for the RDV targets 
were brightness and compactness of the sources \citep{Petrov09}. In this paper, we present results 
of our imaging and analysis of raw archival RDV data for a sample of 370 objects (Fig.~\ref{f:sky}) 
observed during 19 sessions performed within the period 1998--2003. If a source had been observed 
more than once, we selected the epoch at which the dynamic range of the image was its highest. The 
sample is dominated by quasars, with the weak-lined BL Lacs and radio galaxies making up 8.3\% and 
7.8\% of the sample, respectively. We used all 370 sources available from the reduced observations 
for the analysis, because as shown by \cite{LM97}, the most effective approach to studying 
Doppler-boosted sources such as AGNs is based on the investigation of samples with large number of 
objects. The selection effect may otherwise lead to incorrect conclusions about AGN properties as 
a class of radio sources. Chronologically, the RDV project was one of the first among other large 
VLBI imaging surveys covering altogether the frequency range from 2~GHz to 86~GHz and including 
the VLBA Calibrator Survey \citep[VCS;][]{vcs1,vcs2,vcs3,vcs4,vcs5,vcs6}, the VLBI Space Observatory 
Program \citep[VSOP;][]{C_band_pre_VSOP,Dodson_VSOPsurvey}, the VLBI Imaging and Polarimetry Survey 
\citep[VIPS;][]{Hel07,Petrov_VIPS}, the Caltech-Jodrell Bank Flat-Spectrum sample \citep[CJF;][]{PTZ03}, 
the complete sample of BL Lacertae objects \citep{Gabuzda_etal00}, the VLBA 2\,cm Survey 
\citep{2cmPaperI}, and the Monitoring Of Jets in Active galactic nuclei with VLBA Experiments 
\citep[MOJAVE;][]{Lister_etal09}, the VLBI Exploration of Radio Astrometry (VERA) and VLBA K-band 
\citep{Petrov_etal07,VLBA_K_band}, the International Celestial Reference Frame 
\citep[ICRF;][]{Lanyi_etal10,Charlot_etal10} and its extensions \citep{icrf-ext2-2004}, and the 
Coordinated Millimetre VLBI Array \citep[CMVA;][]{Lee08} surveys.

First imaging results for early five 24-hour pre-RDV sessions with the participation of only 
the VLBA antennas were presented by \cite{Fey96} and \cite{Fey97,Fey00}. The imaging of some 
other RDV epochs prior to 1998 was also done by \cite{P07}. In this paper, we present and 
analyze our images from 19 experiments. the VLBI maps for the majority of RDV experiments 
including maps presented in this paper are made available 
online\footnote{\tt http://astrogeo.org/images}\footnote{\tt http://rorf.usno.navy.mil/rrfid.shtml}\footnote{\tt http://www.obs.u-bordeaux1.fr/BVID}.
The astrometric suitability of 80\% of the ICRF sources was investigated by \cite{Charlot08_RDV}.
The RDV observations proved to be successful for spectral index studies including the opacity 
effect detections reported by \cite{Kovalev_cs_2008} and later confirmed by \cite{Sokolovsky_cs_2011}. 
Another application is the measurement of the multi-epoch 8.6~GHz apparent speeds of the jet components.
The initial kinematics results for 54 sources based on 8.6~GHz VLBI images in Radio Reference Frame 
Image Database from the first five years (1994--1998) of data were obtained by \cite{P07}. Extending the period of 
observations up to ten years and enlarging a number of epochs up to 50 will both increase the 
accuracy of the apparent speed measurements and allow us to investigate possible jet acceleration 
\citep{RDV_10yr}. 

The observational material of the RDV project has a number of advantages compared with the data 
from many large VLBA experiments cited above: (i) improved coverage of {\it uv}-plane; (ii) 
simultaneous dual-frequency mode; (iii) high time-sampling frequency (nearly every two months) for a 
subsample of about 60 sources. This third feature permits the study of jet kinematics \citep{P07}, 
while the other two are important for the purposes of this paper, in which we discuss the compactness 
of the sources and both their spectral indices and brightness temperatures, and compare these 
properties for $\gamma$-ray bright and weak sources.

Among the 370 RDV sources presented in this paper, 147 objects have been positionally 
cross-identified with the high-confidence ($>$4$\sigma$) $\gamma$-ray detections associated
with known AGNs, according to the {\it Fermi} Large Area Telescope Second Source Catalog 
\citep[2FGL;][]{2FGL}. These objects are both targets of follow-up observational VLBI sessions 
and prospects for establishing connections between $\gamma$-ray emission and parsec-scale 
jets observed in the radio domain.

The structure of this paper is as follows: in \S~\ref{s:obsproc} we describe the 
observations and data reduction technique, including the calibrating, imaging, and 
model-fitting procedures; in \S~\ref{s:results}, we discuss our results; and our main
conclusions are summarized in \S~\ref{s:summary}. Throughout the paper, the $\Lambda$CDM 
cosmological model with $H_0=70$~km~s$^{-1}$~Mpc$^{-1}$, $\Omega_m=0.3$, and 
$\Omega_\Lambda=0.7$ is adopted. The spectral index $\alpha$ is defined in the convention 
$S\propto\nu^\alpha$. All position angles are given in degrees east of north. We use the 
term ``core'' as the apparent origin of AGN jets that commonly appears as the brightest 
feature in VLBI images of blazars \citep[e.g.,][]{L98,Marscher08}.

\section{Source sample, observational data, and data processing}
\label{s:obsproc}
   
\subsection{Source sample and its completeness}
\label{s:sample}
\begin{figure}
 \resizebox{\hsize}{!}{\includegraphics[angle=-90,clip=true]{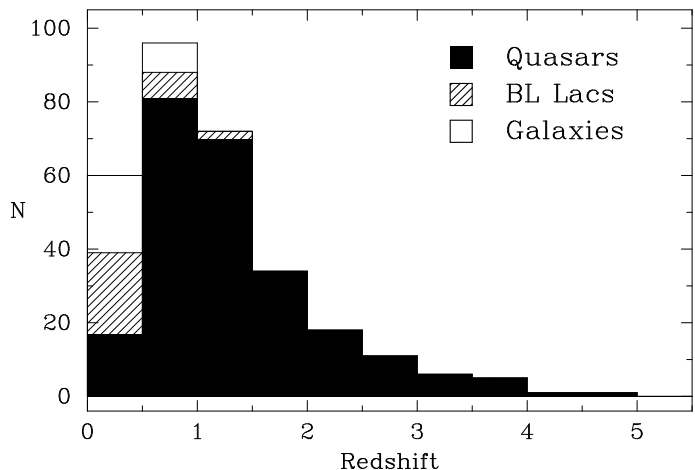}}
 \caption{
          Redshift distribution of 306 sources with the median and maximum values of 0.98 and 4.715,
          respectively.
         }
 \label{f:redshifts}
\end{figure}

The sample of 370 sources includes 251 quasars (67.8\%), 46 BL Lacertae objects (12.4\%), 31 radio
galaxies (8.4\%), and 42 optically unidentified sources (11.4\%). The redshifts are currently
known for 306 objects (83\%). The redshift distribution (Fig.~\ref{f:redshifts}) ranges from
$z_\mathrm{min}=0.00436$ for the galaxy J1230$+$1223 (M87) to $z_\mathrm{max}=4.715$ for the quasar
J1430$+$4204 with a median value of the distribution close to $z=1$. The general characteristics
of the sources such as object name, its alternative name, coordinates of J2000.0 right ascension
and declination, optical class, redshift, and linear scale in parsecs per mas are listed in
Table~\ref{t:general_info}.

\begin{figure}
 \resizebox{\hsize}{!}{\includegraphics[angle=-90,clip=true]{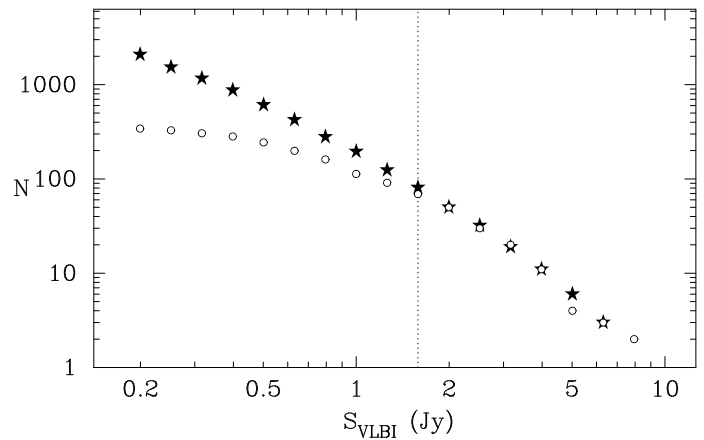}}
 \caption{
          Cumulative $\log N - \log S$ dependence at 8.6~GHz for the studied RDV sample of 370 sources
          (open circles) in comparison with the Radio Fundamental Catalog (stars).
          The dotted line indicates a completeness level of about $1.5$~Jy for the studied sample.
         }
 \label{f:rdv_vs_rfc}
\end{figure}

To investigate the completeness of the studied sample, we determined a cumulative $\log N-\log S$ 
dependence (as shown in Fig.~\ref{f:rdv_vs_rfc}) for the studied RDV subsample and compared it 
with the Radio Fundamental Catalog\footnote{\tt http://astrogeo.org/rfc} (RFC) in the matched 
sky area, with declinations above $-45\degr$. The RFC provides (i) precise positions with 
milli-arcsecond accuracies, (ii) estimates of correlated flux densities at baselines from 
1000~km to 8000~km, and (iii) maps for thousands of compact radio sources produced by the 
analysis of all available VLBI observations under absolute astrometry and geodesy programs. 
The RFC sample constructed from a blind survey is statistically flux density complete down to 
200~mJy \citep{vcs5,vcs6}. We note that the difference of 1--2 sources seen for the brightest 
objects is due to the flux density variability and that the RFC provides the flux densities 
averaged over all available epochs, while the RDV statistics is based on single-epoch measurements. 
As seen from Fig.~\ref{f:rdv_vs_rfc}, the studied RDV subsample is complete down to $\sim$1.5~Jy. 
At lower flux densities, the sample is statistically incomplete but representative of compact 
AGNs owing to the original selection strategy of the RDV project \citep{Petrov09}.

\begin{table*}
\caption{Parameters of observed sources.}
\label{t:general_info}
\begin{center}
\renewcommand{\footnoterule}{}
\begin{tabular}{cccccccc}
\hline
\hline
Source       &  IVS name  &     Right ascension    &        Declination       & Optical &   $z$  &   Scale       & 2FGL   \\
             &            &        J2000.0         &          J2000.0         &  class  &        & pc mas$^{-1}$ & member \\
 (1)         &    (2)     &         (3)            &           (4)            &   (5)   &   (6)  &    (7)        &  (8)   \\
\hline
J0006$-$0623 & 0003$-$066 &  00\,\,06\,\,13.892887 &  $-$06\,\,23\,\,35.33545 &  B      &  0.347 &   4.913       & \ldots \\
J0011$-$2612 & 0008$-$264 &  00\,\,11\,\,01.246737 &  $-$26\,\,12\,\,33.37719 &  Q      &  1.093 &   8.163       & \ldots \\
J0017$+$8135 & 0014$+$813 &  00\,\,17\,\,08.474901 &  $+$81\,\,35\,\,08.13655 &  Q      &  3.387 &   7.407       & \ldots \\
J0019$+$7327 & 0016$+$731 &  00\,\,19\,\,45.786379 &  $+$73\,\,27\,\,30.01753 &  Q      &  1.781 &   8.450       & \ldots \\
J0022$+$0608 & 0019$+$058 &  00\,\,22\,\,32.441214 &  $+$06\,\,08\,\,04.26889 &  B      & \ldots &  \ldots       &      Y \\
J0027$+$5958 & 0024$+$597 &  00\,\,27\,\,03.286191 &  $+$59\,\,58\,\,52.95918 &  U      & \ldots &  \ldots       & \ldots \\
J0035$+$6130 & 0032$+$612 &  00\,\,35\,\,25.310644 &  $+$61\,\,30\,\,30.76129 &  U      & \ldots &  \ldots       & \ldots \\
J0050$-$0929 & 0048$-$097 &  00\,\,50\,\,41.317382 &  $-$09\,\,29\,\,05.21043 &  B      & \ldots &  \ldots       &      Y \\
J0059$+$0006 & 0056$-$001 &  00\,\,59\,\,05.514929 &  $+$00\,\,06\,\,51.62077 &  Q      &  0.717 &   7.215       & \ldots \\
J0102$+$5824 & 0059$+$581 &  01\,\,02\,\,45.762379 &  $+$58\,\,24\,\,11.13659 &  Q      &  0.644$^a$ &   6.902       &      Y \\
\hline
\end{tabular}
\end{center}
Columns are as follows:
(1) J2000.0 IAU name;
(2) IVS name (B1950.0 IAU name or alias);
(3) right ascension (J2000.0) in hours, minutes, and seconds;
(4) declination (J2000.0) in degrees, minutes, and seconds;
(5) optical classification according to \cite{VV12},
    where
    Q is a quasar,
    B is a BL Lacertae object,
    G is an active galaxy, and
    U is unidentified,
(6) redshift as given by \cite{VV12},
    the $a$ flag indicates the redshift from \cite{SE_etal05},
    $b$ -- from \cite{Carilli98},
    $c$ -- from \cite{Drake04},
    $d$ -- from \cite{Afanas_etal03},
    $e$ -- from \cite{Montigny95},
    $f$ -- from \cite{Nilsson08},
    $g$~-- from \cite{Best03},
    $h$ -- from \cite{Hewitt89},
    $k$ -- from \cite{Snellen02a}, and
    $l$ -- from \cite{Snellen02b};
(7) angular scaling conversion in parsecs per milliarcsecond;
(8) membership in the {\it Fermi} LAT Second Source Catalog according to \cite{2FGL}.
The coordinate positions are taken from the Radio Fundamental Catalog (http://astrogeo.org/rfc) 
as derived from an analysis of all VLBI observations made in absolute astrometry and geodesy mode 
\citep{vcs1,vcs2,vcs3,vcs4,vcs5,vcs6}. Table~\ref{t:general_info} is published in its entirety in 
the electronic version of the {\it Astronomy \& Astrophysics}. A portion is shown here for guidance 
regarding its form and content.
\end{table*}

\subsection{Observational data}

We reduced and present here single-epoch imaging results of 19 observational 24-hour sessions 
(RDV11, 13, 15, 18, 21, 24, 27, 29, 30, 31, 32, 33, 34, 35, 36, 37, 38, 39, and 41) carried out 
in a period between October 1998 and September 2003 using a global VLBI array of 18 to 20 stations 
including 10 VLBA antennas and 14 geodetic and EVN radio telescopes (Table~\ref{t:antennas_list})
capable of recording the VLBA modes, using 8 to 10 non-VLBA stations in each session. The dates of 
observations are listed in the first column of Table~\ref{t:gains}. The positions for all antennas 
given in geocentric coordinates are listed in \cite{Petrov09} of Table~3. The observations were 
performed in a standard dual-frequency geodetic mode, registering the signal in right circular 
polarization simultaneously at 2.3~GHz (S-band) and 8.6~GHz (X-band), and using 1-bit sampling. 
Each band was separated into four 8~MHz (4~MHz for RDV11) intermediate frequencies (IFs) giving 
a total observing bandwidth of 32~MHz (16~MHz for RDV11) at each band. The IFs spanned 140~MHz at 
S-band and 490~MHz at X-band (Table~\ref{t:IF_list}). The data were correlated with the VLBA 
correlator at the Array Operation Center in Socorro with 4~s accumulation periods. Most sources 
were observed during six scans of six minutes each, corresponding to a total tracking time of about 
40~min. A few sources were observed in up to 15 scans, making to a total on-source time of about 
60~min. The scans are scheduled over a number of different hour angles to maximize the ($u,v$) 
plane coverage. The median number of independent measurements of visibility function on individual 
interferometer baselines averaged within a scan is about 200. No special calibrators were 
scheduled since all the target sources were compact and bright. More than 92\% of the sources in 
19 processed sessions have VLBI fluxes greater than 200~mJy (see Fig.~\ref{f:flux_hists}, top panels).

Observations were scheduled using the automatic mode of the VLBI scheduling software 
{\it sked} \citep{sked} with some subsequent improvements to maximize mutual visibility, ensure 
adequate sky coverage, minimize slewing time, etc. The detailed description of the 
scheduling of the global geodetic VLBI sessions is discussed by \cite{Petrov09}.

The special feature of these observations is the use of subarrays. The involvement
of the large number of stations located on different continents prevents the antennas 
from being able to simultaneously track the same target source altogether. To 
maximize the on-source time at each station during the whole experiment, the array 
of radio telescopes is divided into several subarrays to ensure that a source is 
visible for every station in a corresponding subarray during a particular scan. The 
number of subarrays ranges from 6 to 8 in the reduced experiments. We note that each 
antenna is included not only in some particular subarray, but in several ones for 
different scans. Thus, the majority of sources are observed with almost all antennas 
participating in a session. The exceptions are the high-declination sources for the 
southern stations and vice versa.

\begin{table}
\begin{minipage}[t]{\columnwidth}
\caption{Antennas participating in the RDV observations.}
\label{t:antennas_list}
\centering
\renewcommand{\footnoterule}{}
\begin{tabular}{l l l c}
\hline\hline
Station       & Location     & Code & Diameter \\
              &              &      &    m     \\
\hline
Algonquin Park& Canada       &  AP  &   46     \\
Brewster      & USA          &  BR  &   25     \\
Fort Davis    & USA          &  FD  &   25     \\
Gilcreek      & USA          &  GC  &   26     \\
Green Bank    & USA          &  GN  &   20     \\
Hancock       & USA          &  HN  &   25     \\
HartRAO       & South Africa &  HH  &   26     \\
Kitt Peak     & USA          &  KP  &   25     \\
Kokkee        & USA          &  KK  &   20     \\
Los Alamos    & USA          &  LA  &   25     \\
Matera        & Italy        &  MA  &   20     \\
Mauna Kea     & USA          &  MK  &   25     \\
Medicina      & Italy        &  MC  &   32     \\
North Liberty & USA          &  NL  &   25     \\
Noto          & Italy        &  NT  &   32     \\
Ny Alesund    & Norway       &  NY  &   20     \\
Onsala        & Sweden       &  ON  &   20     \\
Owens Valley  & USA          &  OV  &   25     \\
Pie Town      & USA          &  PT  &   25     \\
St. Croix     & USA          &  SC  &   25     \\
Concepcion    & Chile        &  TC  &   \phantom{2}6     \\
Tsukuba       & Japan        &  TS  &   32     \\
Westford      & USA          &  WF  &   18     \\
Wettzell      & Germany      &  WZ  &   20     \\
\hline
\end{tabular}
\end{minipage}
\end{table}

\begin{table}
\begin{minipage}[t]{\columnwidth}
\caption{The range of used frequencies.}
\label{t:IF_list}
\centering
\renewcommand{\footnoterule}{}
\begin{tabular}{c c}
\hline\hline
IF &  Frequency range\footnote{S-band IFs were shifted up 12~MHz since the RDV27 session 
(2001.04.09) to get further away from direct broadcast satellite band.}, [MHz] \\
\hline
1  &  $2232.99-2240.99$ \\
2  &  $2262.99-2270.99$ \\
3  &  $2352.99-2360.99$ \\
4  &  $2372.99-2380.99$ \\
5  &  $8405.99-8413.99$ \\
6  &  $8475.99-8483.99$ \\
7  &  $8790.99-8798.99$ \\
8  &  $8895.99-8903.99$ \\
\hline
\end{tabular}
\end{minipage}
\end{table}

The 5~m VLBI antenna MV-3 GGAO near Washington also took part in the observations but 
we did not use the data from this site because of its extremely high noise level 
($\mathrm{SEFD}\gtrsim5\times10^4$~Jy). In addition, GGAO in a number of experiments
was tagged along after the schedule was generated using other stations in a session.
GGAO was added to each scan that the antenna could slew to in time, and the scan 
length for GGAO was set to the maximum of all the stations already participating in 
the scan. This resulted in a significant number of non-detections on the GGAO baselines.

\begin{figure}
 \includegraphics[width=0.39\textwidth,angle=-90,clip=true]{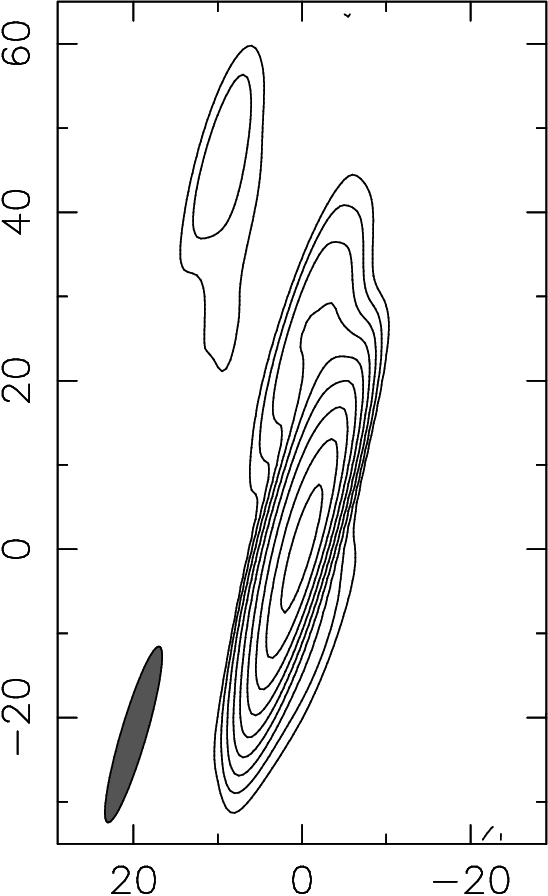}
 \includegraphics[width=0.39\textwidth,angle=-90,clip=true]{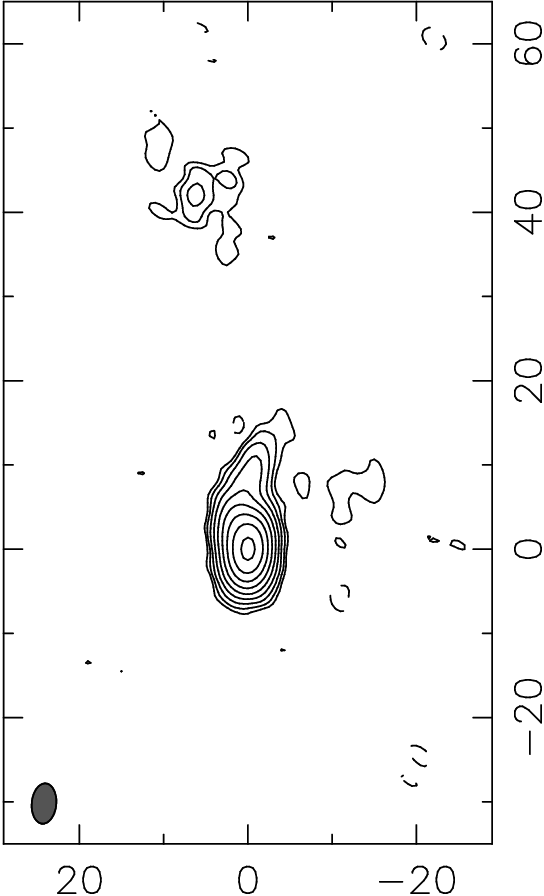}
 \caption{
          Naturally weighted CLEAN images at 2.3~GHz of the low-declination source 
	  J0403-3605 without ({\it left}) and with the data ({\it right}) from 
	  TIGO antenna. The contours are shown at the same levels, starting from 
	  2~mJy~beam$^{-1}$ and increasing by a power of 2. The axes of each image 
	  are given in milliarcseconds. The shaded ellipse in the lower left corner 
	  of each image indicates the full width at half maximum (FWHM) of the 
	  restoring beam.
	 }
 \label{f:TIGO}
\end{figure}

The southern hemisphere antennas, the 26~m radio telescope HartRAO in South Africa 
(latitude of $-26\degr$) and the 6~m TIGO antenna in Concepcion, Chile (latitude of $-36\degr$) 
have participated in 15 out of the 19 analyzed observational sessions. This significantly 
increased the resolution in the north-south direction and allowed us to image low-declination 
sources with almost circular restoring beam. The contribution of data from even such a small 
antenna as TIGO can significantly improve the quality of the final image (Fig.~\ref{f:TIGO}),
if the source is bright ($S>0.5$~Jy).

\begin{figure*}
 \centering
 \includegraphics[width=0.32\textwidth,clip=true,angle=-90]{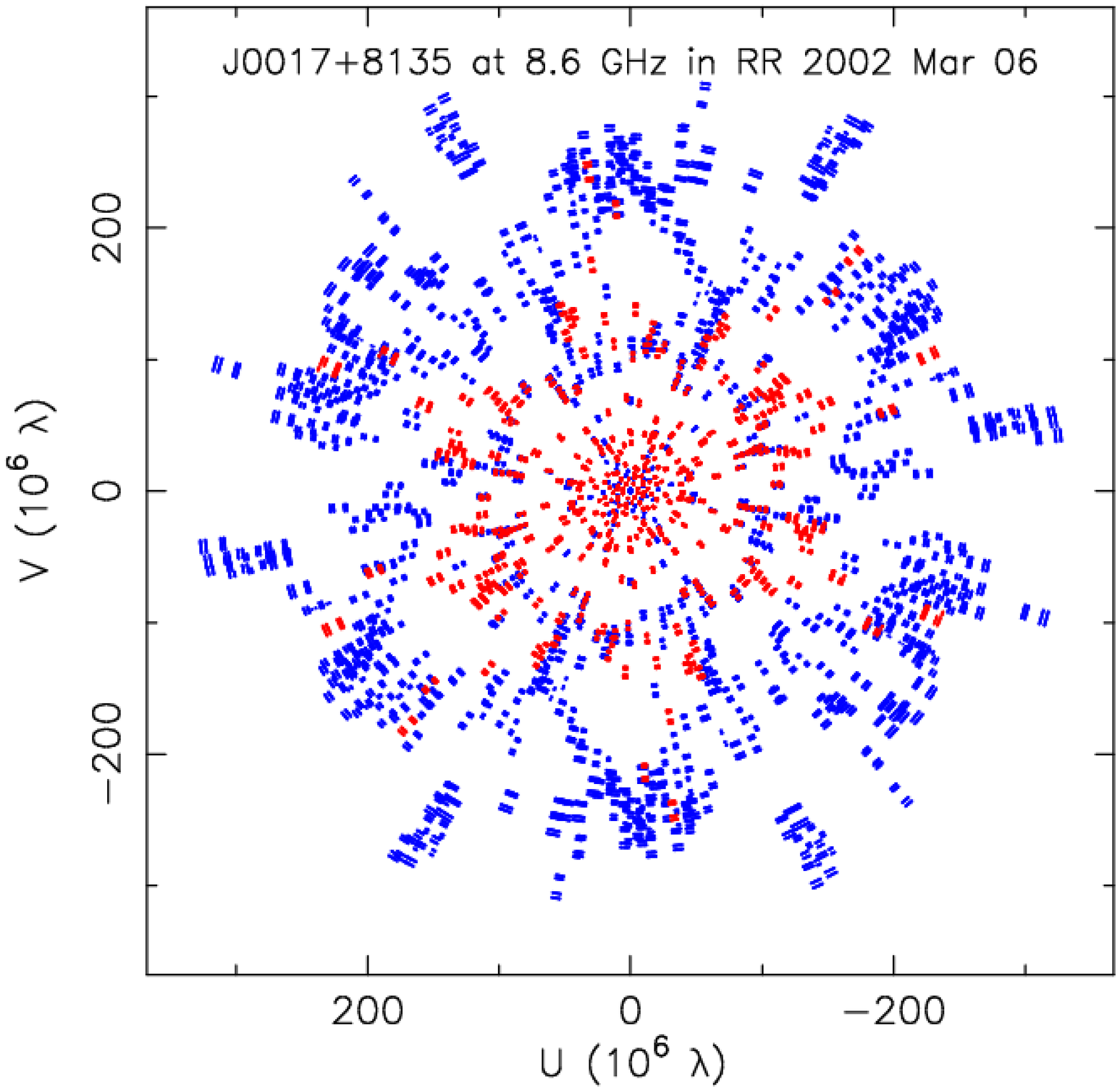}
 \includegraphics[width=0.32\textwidth,clip=true,angle=-90]{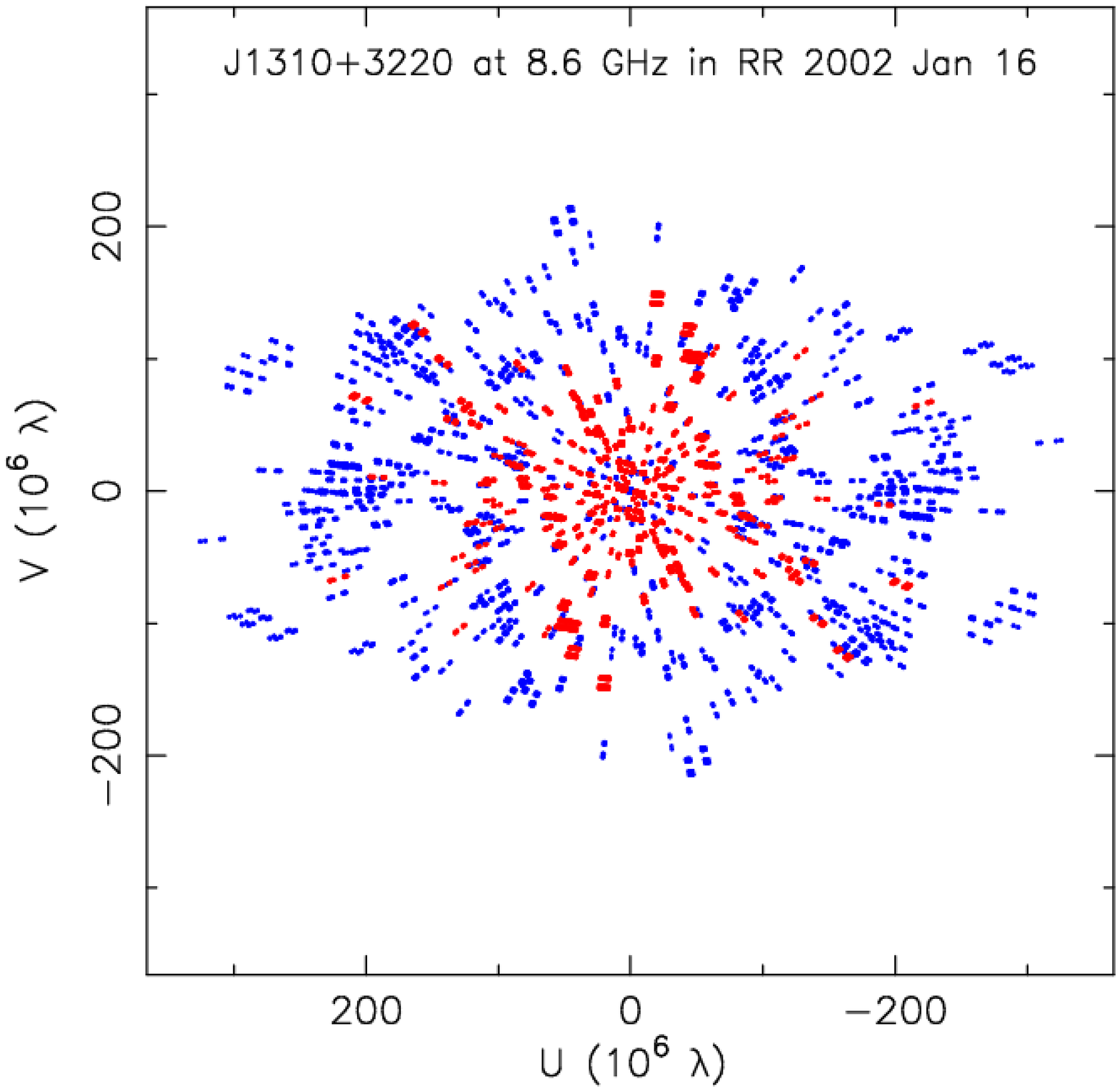}
 \includegraphics[width=0.32\textwidth,clip=true,angle=-90]{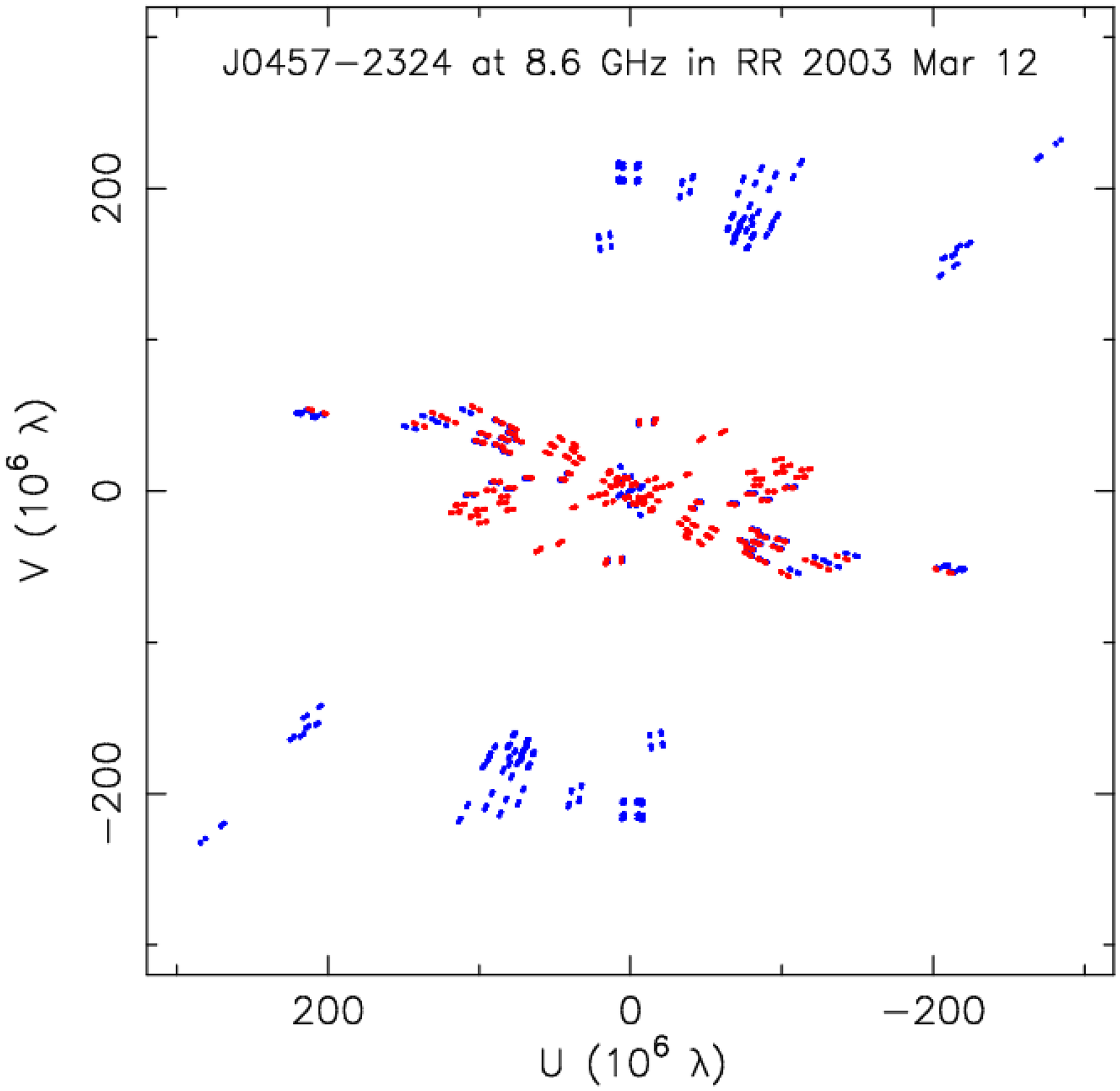}
 \caption{
          Coverages of the ($u,v$) plane at 8.6~GHz for a high-declination (J0017+8135, 
	  {\it left}), medium-declination (J1310+3220, {\it middle}), and low-declination 
	  (J0457$-$2324, {\it right}) source with the baseline projections produced by 
	  the VLBA antennas only (red dots) and the global VLBI array (red + blue dots). 
         }
 \label{f:uv-plane}
\end{figure*}

Increasing the number of antennas in the array from 10 (VLBA) to 18--20 
(global VLBI) allowed us on-average to:
\begin{itemize}
\item increase the number of possible baselines by a factor of $\sim$2.5,
\item increase the angular resolution by a factor of $\sim$1.5,
\item decrease the noise level of the resultant VLBI image by a factor of $\sim$2.
\end{itemize}
The maximum projected baseline of about 12\,300~km between Owens Valley (VLBA) and HartRAO 
that was realized in the RDV experiments is close to the limit for Earth-based VLBI observations. 
In Fig.~\ref{f:uv-plane}, we compare the typical ($u,v$) coverage at 8.6~GHz provided by the 
VLBA and the whole array used for high-, medium-, and low-declination sources. The corresponding 
spatial frequency plane coverages at 2.3~GHz are virtually identical but scaled accordingly.

\begin{table*}
\caption{Observing log with non-VLBA telescope gains. VLBA antennas participated in all sessions.}
\label{t:gains}
\begin{center}
\renewcommand{\footnoterule}{}
\begin{tabular}{ccccccccccccccc}
\hline\hline
Date       & \multicolumn{13}{c}{2.3/8.6~GHz 
                                     DPFU$^a$, [mK Jy$^{-1}$]}\\
\cline{2-15}
           &   AP    &   GC   &   GN  &   HH  &  KK   &  MA   &   MC    &   NY  &   NT   &  ON   & TC  &  TS     &  WF   &  WZ    \\
\hline
1998/10/01 &         & 85/125 & 78/70 &       & 70/41 &       & 140/128 & 24/69 &        & 23/45 &     &         & 43/39 &       \\
1999/03/08 &         &  95/67 & 49/76 & 33/21 & 60/47 &       & 138/152 & 30/77 &        & 24/48 &     &         & 47/40 & 63/68 \\
1999/05/10 &         & 119/74 &       & 36/38 & 60/43 &       & 131/143 & 31/72 &        & 23/34 &     &         & 43/36 & 66/79 \\
1999/12/20 &         &  69/84 & 50/72 & 38/30 & 60/47 &       & 140/146 & 30/77 &        & 23/45 &     & 176/269 & 45/38 & 57/38 \\
2000/05/22 &         &  88/74 &       & 36/24 & 67/39 & 27/45 & 128/118 & 34/73 &        &       &     & 223/271 & 46/37 & 71/80 \\
2000/12/04 &         &  82/94 &       & 36/22 & 65/36 & 47/49 & 180/134 & 32/76 &        &       &     & 205/269 & 44/39 & 68/82 \\
2001/04/09 &         &  95/80 &       & 43/26 & 63/35 & 68/51 & 120/132 & 32/69 &        &       &     & 214/242 & 47/38 & 68/81 \\
2001/07/05 &         &  96/69 &       & 34/20 & 39/38 &       & 126/124 & 31/67 &        &       &     & 207/207 & 54/34 & 62/72 \\
2001/10/29 &         &        &       & 36/15 & 39/41 & 54/40 & 119/114 &       &        &       &     & 154/163 & 60/35 & 65/72 \\
2002/01/16 &         &  98/74 &       &       & 42/41 &       & 139/101 & 46/61 &        & 39/32 &     & 211/236 & 46/33 & 61/53 \\
2002/03/06 &         &  87/81 &       &       & 41/44 &       & 132/116 &       & --/127 & 36/36 &     & 204/203 & 45/34 & 61/71 \\
2002/05/08 & 300/188 &  91/76 &       & 37/16 & 67/42 & 66/42 & 134/113 &       &        &       &     &         & 43/34 & 66/78 \\
2002/07/24 &         &  63/67 &       &       & 71/42 & 51/44 &  --/118 & 61/66 &        & 28/35 & 5/8 &         & 53/34 & 62/69 \\
2002/09/25 &         &  98/71 &       &       & 72/42 & 60/52 &  129/83 &       &        & 24/42 & 5/8 & 309/122 & 56/34 & 71/70 \\
2002/12/11 &         &  98/66 &       &       & 59/40 & 32/41 & 148/133 & 62/67 &        & 31/30 & 4/6 &         & 53/34 & 65/69 \\
2003/03/12 &         &        &       &       & 53/38 & 20/45 & 140/114 &       &        & 28/34 & 5/8 & 227/335 & 77/33 & 62/72 \\
2003/05/07 &         &        &       &       & 53/44 & 29/45 & 129/126 &       &        & 30/30 & 4/8 & 261/212 & 61/36 & 64/81 \\
2003/06/18 &         &        &       &       & 57/42 & --/47 &         &       &        & 31/39 & 6/9 & 220/183 & 66/31 & --/78 \\
2003/09/17 &         &  86/74 &       &       & 61/46 &       &         & 26/54 &        & 20/36 &     & 175/171 & 27/37 & 75/95 \\
\hline
\end{tabular}
\end{center}
$^a$ Efficiency (degrees per flux unit) of a radio telescope in right circular polarization in zenith direction. 
In all cases, the gain curve is assumed to be flat.
\end{table*}

\subsection{Data processing}
Initial calibration was performed with the NRAO Astronomical Image Processing 
System \citep[AIPS;][]{aips} by applying techniques adopted for subarrayed data sets. 
The individual IFs were processed separately throughout the data reduction, and were 
averaged together when making the final images. The antenna gains and system 
temperatures measured during the sessions were used for the amplitude calibration. 
For some of the non-VLBA stations, there was no direct information about the 
antenna's gain at the epoch of an experiment. In these cases, we estimated the degrees 
per flux unit (DPFU) parameter (gain) for each band using the information provided by 
the staff of the observatories about the SEFD measured on a date closest to the epoch 
of observations (to within several days) and the system temperature averaged over the 
current session. Global gain correction factors for each station for each IF were 
derived from the statistical results of self-calibration for all the sources within 
an experiment taking into account the well-measured VLBA antenna gains. If the median 
value of the gain correction distribution deviated by more that 10\% from 1, the whole 
calibration procedure was performed again, where we applied the corresponding corrections 
by running the AIPS task CLCOR prior to the stage of phase calibration. We estimated
the accuracy of amplitude calibration to be at the level of $\sim$10\%. The non-VLBA 
station sensitivities obtained from the log-files and then defined more accurately 
from the gain correction technique are listed in Table~\ref{t:gains}.

Because the phase calibration tones were available for the VLBA stations only (non-VLBA
stations have no phase-cal detectors), the phase corrections for station-based 
residual delays and delay rates were found and applied using the AIPS task FRING in 
two steps. First, the manual fringe fitting was run on a short interval of either one 
or two minutes on a bright compact source to determine the relative instrumental phase 
and residual group delay for each individual IF. As the array of radio telescopes is 
divided into several subarrays, we had to run the procedure on a few different scans to 
find the solution for each antenna. Secondly, the global fringe fitting was run by 
specifying a 4~min solution interval, a point-like source model, and a signal-to-noise 
ratio (S/N) cutoff of 5 to omit noisy solutions. The fraction of inadequate solutions 
was typically at a level of (8--10)\% at X-band and (10--13)\% at S-band.

Amplitude bandpass calibration was not applied because it had no significant effect 
on the dynamic range of our final images. We used the AIPS task SPLIT to write the fully 
calibrated data for each individual source to separate X- and S-band FITS-files averaging 
data in frequency within each of four IFs at each band, and making no averaging in time. 
First, phase self-calibration with a point-source model was done using the AIPS task CALIB, 
which provides: (i) the possibility of finding the first solution using all IFs, (ii) automatic 
flagging the data with low S/N ratio. 

CLEANing \citep{CLEANref}, phase and amplitude self-calibration \citep{Jennison58,Twiss_etal60}, 
and hybrid imaging \citep{Readhead_etal80,Schwab80,CW81} were performed in the Caltech DIFMAP 
\citep{difmap} package using an automated approach suggested by G.~Taylor \citep{difmap-script}, 
which we adapted to the RDV data set. In all cases, a point-source model was used as an initial 
model for the iterative procedure. Final maps were produced by applying a natural weighting of 
the visibility function. The spanned bandwidth of the four IFs in each band is relatively small 
(6\% of fractional bandwidth in both bands), thus no spectral correction technique was applied. 
In the model fitting also performed in DIFMAP, we used a minimum number of circular (and in 
some cases elliptical) Gaussian components that after being convolved with the restoring beam, 
adequately reproduce the constructed brightness distribution. 

\section{Results}
\label{s:results}

\begin{table*}
\caption{Parameters of VLBI maps.}
\label{t:map_parameters}
\begin{center}
\renewcommand{\footnoterule}{}
\begin{tabular}{c|ccccccr|ccccccr}
\hline 
\hline
Source &   \multicolumn{7}{c|}{2.3~GHz} & \multicolumn{7}{c}{8.6~GHz} \\
\cline{2-15}
       & peak         & clev          & $S_\mathrm{VLBI} $ & $S_\mathrm{unres}$ & $B_\mathrm{min}$ & $B_\mathrm{maj}$ & $B_\mathrm{PA}$ & peak          & clev          & $S_\mathrm{VLBI}$ & $S_\mathrm{unres}$ & $B_\mathrm{min}$ & $B_\mathrm{maj}$ & $B_\mathrm{PA}$ \\
       & Jy bm$^{-1}$ & mJy bm$^{-1}$ & Jy                 & Jy              & mas              & mas              & deg             & Jy bm$^{-1}$  & mJy bm$^{-1}$ & Jy                & Jy              & mas              & mas              & deg             \\
 (1)   & (2)          & (3)           & (4)                & (5)             & (6)              & (7)              & (8)             & (9)           & (10)          & (11)              & (12)            & (13)             & (14)             & (15)            \\
\hline
J0006$-$0623 & 1.33 & 3.12 & 2.33 & 0.66 & 3.42 & 8.03 &     4 & 1.41 & 2.92 & 2.07 & 1.18 & 0.91 &  2.23 &     4 \\
J0011$-$2612 & 0.39 & 1.90 & 0.41 & 0.36 & 2.06 & 5.56 &  $-$5 & 0.31 & 1.60 & 0.34 & 0.27 & 0.56 &  2.03 & $-$10 \\
J0017$+$8135 & 0.52 & 0.96 & 0.76 & 0.40 & 2.55 & 2.71 &    54 & 0.49 & 0.76 & 0.75 & 0.29 & 0.68 &  0.69 &     7 \\
J0019$+$7327 & 0.57 & 1.20 & 0.76 & 0.35 & 2.51 & 2.86 &    20 & 0.15 & 1.00 & 0.34 & 0.13 & 0.68 &  0.74 &    19 \\
J0022$+$0608 & 0.42 & 0.79 & 0.45 & 0.40 & 3.12 & 6.50 &  $-$2 & 0.41 & 0.58 & 0.44 & 0.35 & 0.84 &  1.76 &  $-$1 \\
J0027$+$5958 & 0.07 & 0.90 & 0.08 & 0.07 & 2.81 & 3.63 & $-$23 & 0.06 & 0.90 & 0.07 &\ldots& 0.65 &  1.02 & $-$24 \\
J0035$+$6130 & 0.07 & 2.06 & 0.11 & 0.09 & 4.01 & 5.00 &    19 & 0.08 & 1.94 & 0.11 & 0.07 & 1.07 &  1.34 &     2 \\
J0050$-$0929 & 0.53 & 1.93 & 0.58 & 0.45 & 2.97 & 9.62 &  $-$8 & 0.70 & 1.65 & 0.75 & 0.61 & 0.80 &  2.62 &  $-$7 \\
J0059$+$0006 & 0.42 & 5.80 & 1.17 & 0.25 & 4.38 & 7.68 &  $-$4 & 0.16 & 5.40 & 0.43 & 0.06 & 1.14 &  2.18 &  $-$8 \\
J0102$+$5824 & 0.61 & 0.92 & 0.81 & 0.42 & 2.63 & 3.25 &    32 & 1.55 & 0.88 & 1.68 & 1.46 & 0.67 &  0.87 &    35 \\
\hline
\end{tabular}
\end{center}
Columns are as follows:
(1) source name (J2000.0),
(2) peak flux density in image,
(3) lowest contour in image,
(4) total flux density from VLBI map,
(5) flux density from the most compact component on map,
(6) FWHM minor axis of restoring beam,
(7) FWHM major axis of restoring beam,
(8) position angle of major axis of restoring beam,
(9)--(15) the same as (2)--(8), respectively.
Table~\ref{t:map_parameters} is published in its entirety in the electronic version of the {\it Astronomy \& Astrophysics}.
A portion is shown here for guidance regarding its form and content.
\end{table*}

\subsection{Calibrated VLBI images}
\begin{figure*}[p]
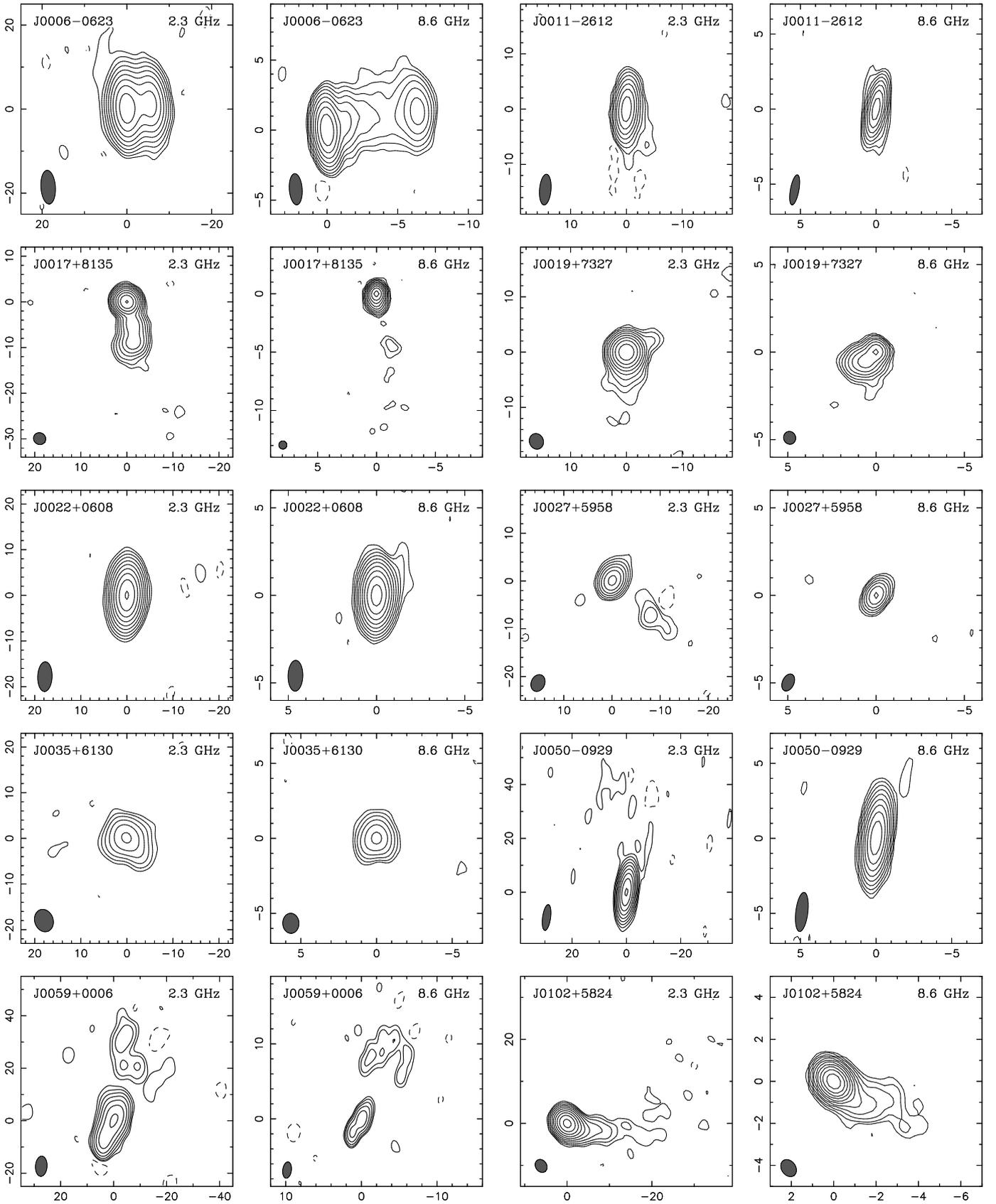

\centering
\includegraphics[width=45mm, angle=-90, trim=-15 -15 -15 -15]{FIGS/MAPS/J0006-0623_S_2002_03_06_pus_map.ps}
\includegraphics[width=45mm, angle=-90, trim=-15 -15 -15 -15]{FIGS/MAPS/J0006-0623_X_2002_03_06_pus_map.ps}
\includegraphics[width=45mm, angle=-90, trim=-15 -15 -15 -15]{FIGS/MAPS/J0011-2612_S_1999_03_08_pus_map.ps}
\includegraphics[width=45mm, angle=-90, trim=-15 -15 -15 -15]{FIGS/MAPS/J0011-2612_X_1999_03_08_pus_map.ps}
\includegraphics[width=45mm, angle=-90, trim=-15 -15 -15 -15]{FIGS/MAPS/J0017+8135_S_2002_03_06_pus_map.ps}
\includegraphics[width=45mm, angle=-90, trim=-15 -15 -15 -15]{FIGS/MAPS/J0017+8135_X_2002_03_06_pus_map.ps}
\includegraphics[width=45mm, angle=-90, trim=-15 -15 -15 -15]{FIGS/MAPS/J0019+7327_S_1999_03_08_pus_map.ps}
\includegraphics[width=45mm, angle=-90, trim=-15 -15 -15 -15]{FIGS/MAPS/J0019+7327_X_1999_03_08_pus_map.ps}
\includegraphics[width=45mm, angle=-90, trim=-15 -15 -15 -15]{FIGS/MAPS/J0022+0608_S_2002_01_16_pus_map.ps}
\includegraphics[width=45mm, angle=-90, trim=-15 -15 -15 -15]{FIGS/MAPS/J0022+0608_X_2002_01_16_pus_map.ps}
\includegraphics[width=45mm, angle=-90, trim=-15 -15 -15 -15]{FIGS/MAPS/J0027+5958_S_2001_10_29_pus_map.ps}
\includegraphics[width=45mm, angle=-90, trim=-15 -15 -15 -15]{FIGS/MAPS/J0027+5958_X_2001_10_29_pus_map.ps}
\includegraphics[width=45mm, angle=-90, trim=-15 -15 -15 -15]{FIGS/MAPS/J0035+6130_S_2002_12_11_pus_map.ps}
\includegraphics[width=45mm, angle=-90, trim=-15 -15 -15 -15]{FIGS/MAPS/J0035+6130_X_2002_12_11_pus_map.ps}
\includegraphics[width=45mm, angle=-90, trim=-15 -15 -15 -15]{FIGS/MAPS/J0050-0929_S_2002_03_06_pus_map.ps}
\includegraphics[width=45mm, angle=-90, trim=-15 -15 -15 -15]{FIGS/MAPS/J0050-0929_X_2002_03_06_pus_map.ps}
\includegraphics[width=45mm, angle=-90, trim=-15 -15 -15 -15]{FIGS/MAPS/J0059+0006_S_1999_05_10_pus_map.ps}
\includegraphics[width=45mm, angle=-90, trim=-15 -15 -15 -15]{FIGS/MAPS/J0059+0006_X_1999_05_10_pus_map.ps}
\includegraphics[width=45mm, angle=-90, trim=-15 -15 -15 -15]{FIGS/MAPS/J0102+5824_S_2002_01_16_pus_map.ps}
\includegraphics[width=45mm, angle=-90, trim=-15 -15 -15 -15]{FIGS/MAPS/J0102+5824_X_2002_01_16_pus_map.ps}
\caption{
 One-epoch 2.3~GHz and 8.6 GHz VLBI images of 370 active galactic nuclei. Image parameters are
 listed in Table~\ref{t:map_parameters}. Gaussian models fitted to the visibility data at each 
 frequency are listed in Table~\ref{t:models}. The scale of each image is in milliarcseconds.
 The shaded ellipse in the lower left corner indicates the FWHM of the restoring beam (with 
 natural weighting). The contours are plotted at increasing factors of 2 times the lowest 
 contour drawn at $3\sigma$ level. A single negative contour equal to the bottom positive 
 contour is also shown. A cell size of 0.2 and 0.5 mas per pixel was used for 8.6 and 2.3~GHz 
 maps, respectively. Fig.~\ref{f:maps} is available in its entirety in the electronic version 
 of the {\it Astronomy \& Astrophysics}.
}
\label{f:maps}
\end{figure*}

Final, naturally weighted VLBI images at 2.3 and 8.6~GHz for 370 sources are shown in Fig.~\ref{f:maps}.
The dynamic range of the images determined as a ratio of the peak flux density to the rms noise level 
ranges from 66 to 7042 with a median of $\sim$1200 at X-band, and ranges from 106 to 4789 with a median 
of $\sim$1000 at S-band. The typical rms noise level is $\sim$0.4~mJy~beam$^{-1}$ at X-band and 
$\sim$0.5~mJy~beam$^{-1}$ at S-band. 

In Table~\ref{t:map_parameters}, we summarize the VLBI map parameters: (1) source name in J2000.0 
notation; for X-band: (2) peak flux density in Jy~beam$^{-1}$; (3) lowest contour level in 
mJy~beam$^{-1}$; (4) flux density integrated over entire map in Jy; (5) flux density of the most 
compact fraction in Jy as measured from the longest baselines; (6) major axis of the full width at
half maximum (FWHM) of the restoring beam in milliarcsec; (7) minor axis of FWHM of restoring beam 
in milliarcsec; (8) position angle of major axis of restoring beam in degrees; columns (9)--(15) 
are the same as (2)--(8) but for S-band. Flux densities shown in columns (4) and (11) were calculated 
as a sum of all CLEAN components over VLBI image. Columns (5) and (12) give the flux densities of 
the unresolved fractions $S_\mathrm{unres}$ and are estimated as the median of the visibility 
function amplitudes measured at spatial $(u,\,v)$ radius $r_{uv}=(u^2+v^2)^{1/2}>270$~M$\lambda$ 
for X-band and $r_{uv}>72$~M$\lambda$ for S-band, corresponding to the range between 0.8 and 1.0 
of the maximum projected spacing. For 15 sources at X-band and 6 sources at S-band, no estimates 
of unresolved flux density are given because no data were available on the baselines used for the 
calculations.

In addition to Table~\ref{t:map_parameters}, all calibrated data including flexible image transport 
system (FITS) files of naturally weighted CLEAN images, FITS files with ($u,v$) data sets, images in 
postscript and GIF-formats, plots of correlated flux density versus baseline length projection, and 
values of total VLBI flux density for all the sources at all available epochs listed in Table~\ref{t:gains} 
are accessible online through the HTML interface.\footnote{See \tt http://vlbi.crao.crimea.ua/SX}

\begin{figure}
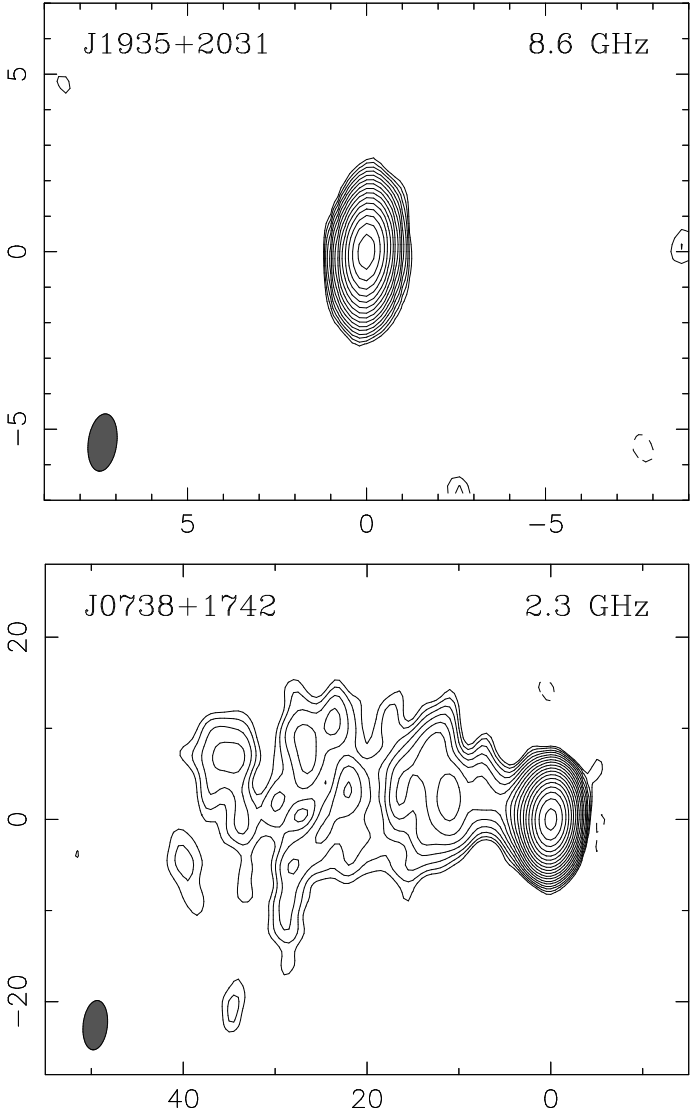

   \resizebox{\hsize}
{!}{\includegraphics[clip=true,angle=-90]{FIGS/fig07_map_compact_pdl.eps}}
   \resizebox{\hsize}{!}{\includegraphics[clip=true,angle=-90,trim=-20 0 0 0]{FIGS/fig07_map_resolved_pdl.eps}}
   \caption{
            Naturally weighted CLEAN images of J1935+2031 at 8.6~GHz ({\it top}, dynamic range of 570) and J0738+1742 at 
            2.3~GHz ({\it bottom}, dynamic range of 1750) as examples of highly compact and jet-resolved sources, respectively. 
	    The axes of each image are given in milliarcseconds. The shaded ellipse in the lower left corner 
	    indicates the FWHM of the restoring beam.}
              \label{f:map_examples}%
    \end{figure}

\begin{figure}
\resizebox{\hsize}{!}{\includegraphics[width=0.36\textwidth,clip=true,angle=-90]{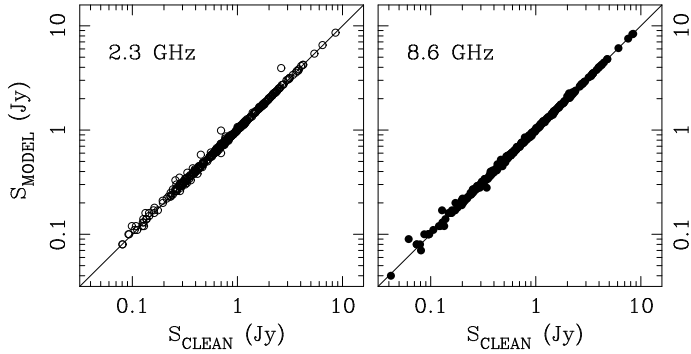}}
   \caption{
            Total flux density from Gaussian fits versus total flux density from VLBI CLEAN image for the data
            at 2.3~GHz (left) and 8.6~GHz (right).}
              \label{clean_vs_model}%
    \end{figure}

\subsection{Parsec-scale structure modeling}
The observed sources have very complex three-dimensional structures on parsec and dekaparsec scales. 
However, since they are located at cosmological distances they provide us with the possibility of 
investigating only their two-dimensional brightness distributions in the plane of the sky, which 
can be modeled by a limited number of Gaussian components. The number of components is usually 
determined by reconciling two opposing conditions: (i) it should be minimal; (ii) all the components 
are convolved with the restoring beam, adequately reproducing the general structure of the obtained 
brightness distribution. The typical morphology of the sources is represented by a one-sided core-jet 
structure, which is the result of strong selection effects and the Doppler boosting of jet emission 
\citep{Cohen_beaming}. Only 11 sources (3\% of the sample) have two-sided structures detected in the 
S/X images. There are also extreme cases when (i) a source is highly compact without any indications 
of a jet, presumably because of lack of spatial resolution and/or sensitivity of the interferometer 
at the observed frequency; (ii) a source has a prominent jet that is resolved even in a transverse 
direction. In Fig.~\ref{f:map_examples}, we plot the images of J1935+2031 and J0738+1742 as examples.

Structure modeling of all sources was performed with the procedure {\it modelfit} in DIFMAP package 
by fitting several circular Gaussian components to the self-calibrated visibility data and minimizing 
$\chi^2$ in the spatial frequency plane. Elliptical Gaussian components were adopted when the use of 
circular Gaussian component resulted in the appearance of axisymmetric noise on the residual map near 
the modeled component. The model fits listed in Table~\ref{t:models} provide flux densities, positions, 
and sizes for the bright and distinct jet features. All the positions are given with respect to the core
component. The quality of model fitting is generally quite good as can be seen in Fig.~\ref{clean_vs_model}, 
where we plot the total flux density of the Gaussian components of the fit and CLEANed flux density 
calculated as a sum of all CLEAN components from the image.

\begin{table}
\caption{Source models.}
\label{t:models}
\begin{center}
\renewcommand{\footnoterule}{}
\begin{tabular}{ccrccccc}
\hline
\hline
Source       &  B  & $S$~~\, &  $r$   & $\varphi$ &   Maj.  &  Ratio  & P.A.  \\
             &     &  mJy~   &  mas   &  deg      &   mas   &         & deg   \\
 (1)         & (2) &  (3)~\, &  (4)   &  (5)      &   (6)   &   (7)   & (8)   \\
\hline
J0006$-$0623 &  S  &  1527   & \ldots &  \ldots   &   1.87  &   0.93  & $-$37 \\
             &     &    46   &   2.96 &   $-$55   & \ldots  &   1.00  &\ldots \\
             &     &   772   &   5.56 &   $-$77   &   2.79  &   1.00  &\ldots \\
\cline{3-8}
             &  X  &  1551   & \ldots &  \ldots   &   0.81  &   0.34  &     7 \\
             &     &   112   &   1.26 &   $-$68   &   0.50  &   1.00  &\ldots \\
             &     &   115   &   2.23 &   $-$65   &   1.58  &   1.00  &\ldots \\
             &     &   128   &   5.34 &   $-$76   &   2.77  &   1.00  &\ldots \\
             &     &   180   &   6.58 &   $-$78   &   0.87  &   1.00  &\ldots \\
\hline
J0011$-$2612 &  S  &   401   & \ldots &  \ldots   &   0.57  &   1.00  &\ldots \\
\cline{3-8}
             &  X  &   327   & \ldots &  \ldots   &   0.26  &   1.00  &\ldots \\
\hline
\end{tabular}
\end{center}
Columns are as follows:
(1) source name (J2000.0),
(2) frequency band,
(3) fitted Gaussian flux density,
(4) position offset from the core component,
(5) position angle of the component with respect to the core component,
(6) FWHM major axis of the fitted Gaussian,
(7) axial ratio of the fitted Gaussian,
(8) major axis position angle of the fitted Gaussian.
Table~\ref{t:models} is published in its entirety in the electronic version of the {\it Astronomy \& Astrophysics}.
A portion is shown here for guidance regarding its form and content.
\end{table}

\begin{figure}[t!]
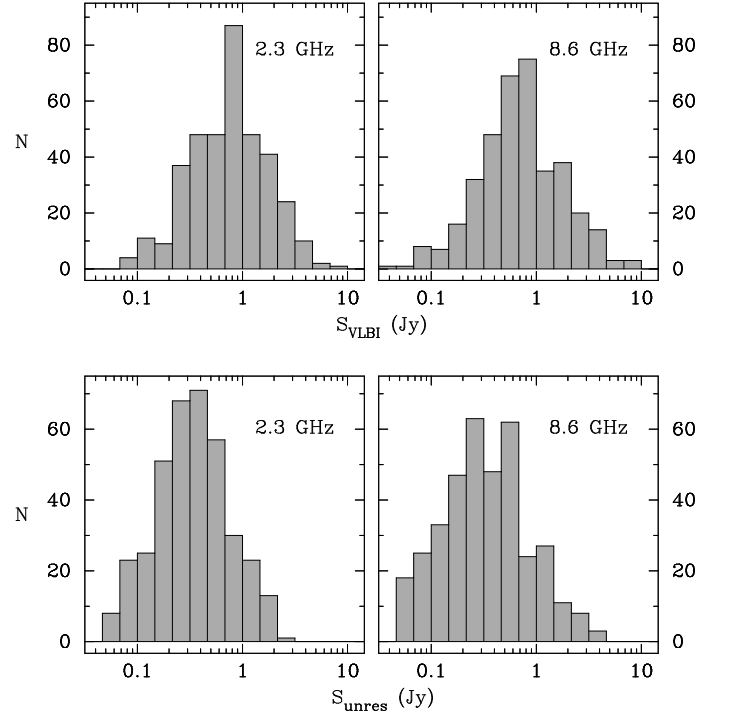

 \resizebox{\hsize}{!}{\includegraphics[clip=true,angle=-90]{FIGS/fig09_total_flux_gray.eps}}
 \resizebox{\hsize}{!}{\includegraphics[clip=true,angle=-90,trim=-20 0 0 0]{FIGS/fig09_unres_flux_gray.eps}}
 \caption{
          Distributions of the VLBI flux density $S_\mathrm{VLBI}$ ({\it top panels}) and 
          the flux density of the most compact component $S_\mathrm{unres}$ ({\it bottom panels}).}
 \label{f:flux_hists}%
\end{figure}

\subsection{Compactness}

One of the main goals of the RDV program was to construct the radio reference frame. The sources 
included in the observing program thus have to lie at cosmological distances, be bright and compact, 
i.e., to radiate the emission from the extremely limited sky area. The most suitable objects for 
achieving these goals are active galactic nuclei, which easily satisfy the aforementioned criteria. 
The compactness of blazars on arcsecond scales derived as $S_\mathrm{VLBI}/S_\mathrm{tot}$, where 
$S_\mathrm{tot}$ is the single-dish flux densities and $S_\mathrm{VLBI}$ is the integrated flux density 
from a VLBI image, is at a very high level of $\sim$0.9 at 2~GHz and 8~GHz as initially reported by 
\cite{PopovKovalev99} for a small sample of 20 sources and later confirmed by \cite{2cmPaperIV} for a 
complete sample of 250 flat-spectrum AGNs observed with the VLBA at 15~GHz. This implies that nearly 
all of the emission from these sources is generated on the milliarcsecond scales, probed by VLBI 
observations, and that the contribution from extended emission on kiloparsec scales is typically 
negligible. 

\begin{figure}[t!]
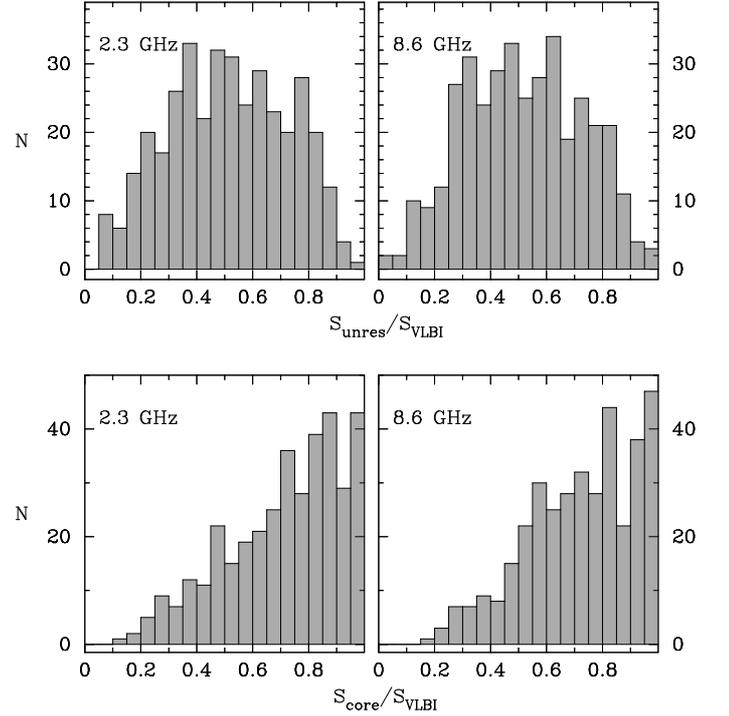

 \resizebox{\hsize}{!}{\includegraphics[clip=true,angle=-90]{FIGS/fig10_comp_total_gray.eps}}
 \resizebox{\hsize}{!}{\includegraphics[clip=true,angle=-90,trim=-20 0 0 0]{FIGS/fig10_comp_core_gray.eps}}
 \caption{
          Distributions of the compactness $S_{\mathrm{unres}}/S_{\mathrm{VLBI}}$ ({\it top panels}) 
          and the VLBI core dominance $S_{\mathrm{core}}/S_{\mathrm{VLBI}}$ ({\it bottom panels}).
	 }
 \label{f:compactness}%
\end{figure}

The compactness of the sources from our sample on milliarcsecond scales, determined as a ratio of 
$S_\mathrm{unres}$ to $S_\mathrm{VLBI}$, distributions of which are presented in Fig.~\ref{f:flux_hists}, 
is statistically lower (Fig.~\ref{f:compactness}, top panels) with median values of 0.51 
at 2.3 and 8.6~GHz (Table~\ref{t:statistics}). According to both K-S and T-test, the distributions 
are indistinguishable. At higher frequencies, the VLBI compactness is also at the similar level. 
Thus, 68\% of the sample of 250 extragalactic radio sources observed at 15~GHz have compactnesses
greater than 0.5 \citep{2cmPaperIV}. The median value of compactness at 86~GHz is about 0.5 as 
reported by \cite{Lee08}.

We note that the overwhelming majority of the sources are also strongly core dominated 
(Fig.~\ref{f:compactness}, bottom panels). In 50\% of sources, the emission from the core region 
accounts for more than 75\% of the total VLBI flux. The information on $S_\mathrm{VLBI}$, 
$S_\mathrm{unres}$, and compactness will be useful for selecting the most suitable candidates 
(highly-compact bright sources) to be studied with the {\it RadioAstron} space VLBI mission 
\citep{RadioAstron}. At the same time, there is a fraction of the sources that manifest strong 
jets (M87, 3C~120, etc.), which are of particular interest for detailed astrophysical studies 
of their extended outflows on parsec scales.

\begin{figure}
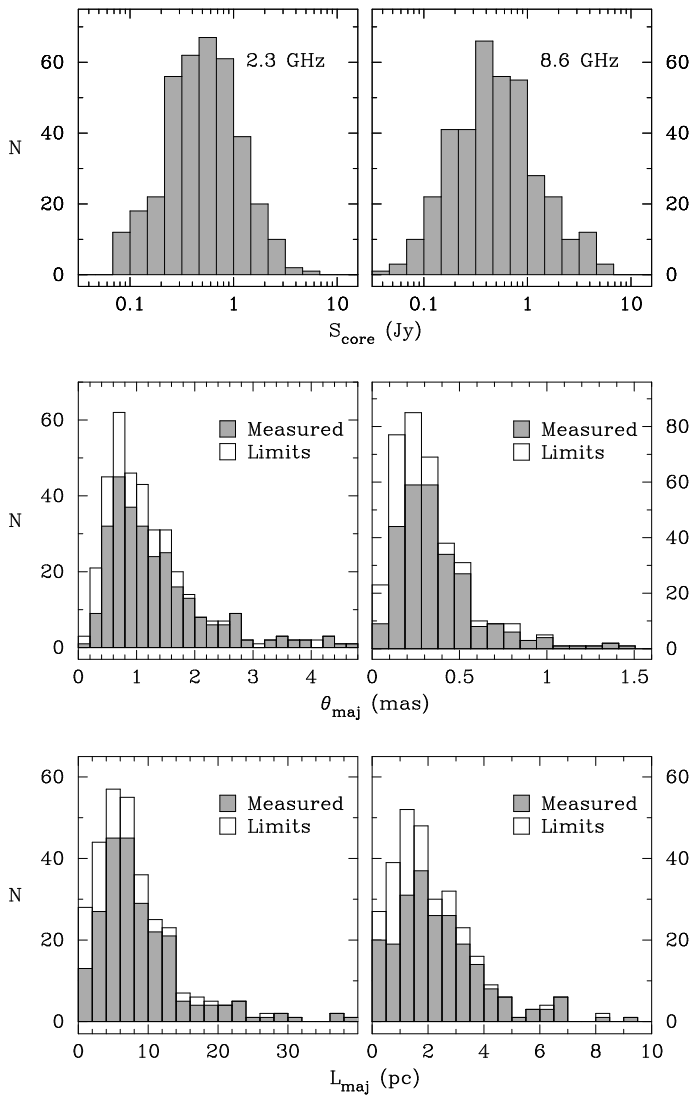

 \resizebox{\hsize}{!}{\includegraphics[clip=true,angle=-90]{FIGS/fig11_core_flux_gray.eps}}
 \resizebox{\hsize}{!}{\includegraphics[clip=true,angle=-90,trim=-20 0 0 0]{FIGS/fig11_core_size_ang_gray.eps}}
 \resizebox{\hsize}{!}{\includegraphics[clip=true,angle=-90,trim=-20 0 0 0]{FIGS/fig11_core_size_lin_gray.eps}}
 \caption{
          Distributions of flux ({\it top}), angular size ({\it middle}), and projected linear size ({\it bottom}) 
          for the core components at 2.3~GHz ({\it left}) and 8.6~GHz ({\it right}). The unshaded boxes 
          represent upper limits for those sources with unresolved core component. For the core components
          fitted with elliptical Gaussians, the major axis is used for the angular core size.
	 }
 \label{f:core_hists}%
\end{figure}

\begin{figure}
 \resizebox{\hsize}{!}{\includegraphics[clip=true,angle=-90]{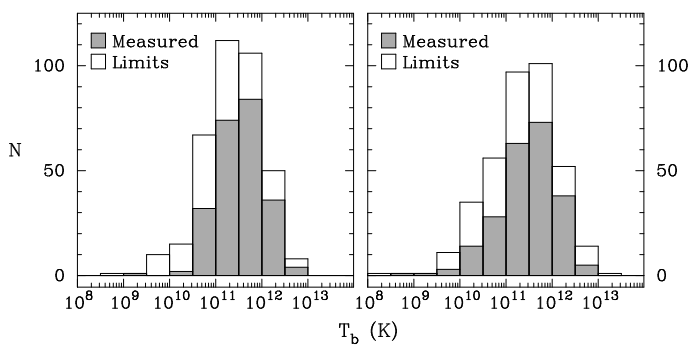}}
 \caption{
          Brightness temperature distribution of VLBI core components in source rest frame at 2.3~GHz 
	  ({\it left}) and 8.6~GHz ({\it right}) with median value of $2.51\times10^{11}$~K and 
	  $2.60\times10^{11}$~K, respectively. The unshaded boxes represent lower limits for those 
	  sources with either unknown redshift or unresolved core component.
	 }
 \label{f:Tb_hists}%
\end{figure}

\subsection{VLBI core properties}

We define VLBI core as a compact, bright emitting region at the narrow end of the jet, where it becomes 
optically thick. Typically, VLBI cores are often unresolved along one or both axes of the restoring beam 
and have a flat radio spectrum. Since the core is usually the brightest feature of the source it is placed 
by the phase self-calibration procedure at the center of the map. In such a case, the core component 
always has a very small but nonzero shift from the phase center because of the close presence of other 
bright components down the jet. Significant offsets of the VLBI core components from the occur phase 
center of an image are rare and happen when one of the jet components is brighter than the core. This 
kind of structure has been found for 18 sources at X-band, and for 27 ones at S-band. 

Model fitting enabled us to measure flux densities and sizes of the core components. The distributions of 
the core flux densities are plotted in Fig.~\ref{f:core_hists} (top panels). Owing to the high compactness 
of the cores, they are often unresolved. We considered the core to be unresolved if the size of a circular 
Gaussian component or at least one of the axes of an elliptical Gaussian component was smaller than the 
respective resolution limit (which is a function of the restoring beam parameters and S/N at the core 
position) calculated following \cite{2cmPaperIV}. The resolution limits were used as upper limits to the 
component sizes. Overall, nearly one-fourth of the cores were unresolved, namely 98 at X-band and 87 at 
S-band. The projected linear sizes could also be calculated for the sources with known redshifts. The 
corresponding distributions of the angular and projected linear sizes of the VLBI core components are 
presented in Fig.~\ref{f:core_hists} (middle and bottom panels), with median values of 0.28~mas (1.90~pc) 
and 1.04~mas (6.75~pc) at X-band and S-band measurements, respectively.

The brightness temperature of the core components in the source rest frame is given by \citep[e.g.,][]{2cmPaperIV}
\begin{equation}
T_\mathrm{b,\ core}=\frac{2\ln2}{\pi k}\frac{S_\mathrm{core}\lambda^2(1+z)}{\theta_\mathrm{maj}\theta_\mathrm{min}},
\end{equation}
where $k$ is the Boltzmann constant, $S_\mathrm{core}$ is the VLBI core flux density, $\theta_\mathrm{maj}$ 
and $\theta_\mathrm{min}$ are the FWHMs of the elliptical Gaussian component along the major and the minor 
axes respectively, $\lambda$ is the wavelength of observation, and $z$ is the redshift. Measuring the flux 
density in Jy, component sizes in mas, and wavelength in cm, we obtained
\begin{equation}
T_\mathrm{b,\ core}=a_{\lambda}\frac{S_\mathrm{core}(1+z)}{\theta_\mathrm{maj}\theta_\mathrm{min}}\,\,\mathrm{K},
\end{equation}
where $a_{\lambda }=1.639\times10^{10}$ for X-band ($\lambda=3.5$~cm) and $a_{\lambda }=2.337\times10^{11}$ 
for S-band ($\lambda=13$~cm). For sources without measured redshifts (see Table~\ref{t:general_info}) we used 
$z=0$ to determine a lower limit to the brightness temperature. Another fraction of sources that also yielded 
lower limits to the $T_\mathrm{b}$ values was a category of objects with unresolved cores. Altogether, about 
40\% of the $T_\mathrm{b}$ values are lower limits. 

The distributions of the brightness temperatures of the VLBI cores at 2.3~GHz and 8.6~GHz shown in 
Fig.~\ref{f:Tb_hists} are broad covering more than four orders of magnitude with the median value of 
$~2.5\times10^{11}$~K at both frequencies, which is close to both the equipartition value of 
$\sim$$5\times10^{11}$~K \citep{R94} and the inverse Compton limit of $\sim$$10^{12}$~K \citep{KPT69}. 
The higher brightness temperatures may be attributed to Doppler boosting 
\citep[e.g.,][]{2cmPaperIV,Homan_Tb_2006,Cohen_beaming,Kellermann_etal07}. The histograms in 
Fig.~\ref{f:Tb_hists} obtained at different frequencies of 2 and 8~GHz, look similar because the maximum 
measurable $T_\mathrm{b}$ is determined by the physical length of the effective projected baseline and 
the core flux density \citep[e.g.,][]{2cmPaperIV,Kovalev_VLBIsurveys09}. Therefore, for VLBI observations 
simultaneously made at different frequencies, we would expect to obtain close $T_\mathrm{b}$ values, 
taking into account the flat radio spectra of the core components in most cases. The VLBI core parameters, 
such as the fitted flux density, sizes, or estimated resolution limits, and derived brightness temperature 
separately for 2.3~GHz and 8.6~GHz are listed in Table~\ref{t:Tb_core}. The statistics of the distributions 
shown in Figs.~\ref{f:flux_hists}--\ref{f:Tb_hists} are summarized in Table~\ref{t:statistics}. The mean 
values of the distributions of $\theta_\mathrm{maj}$, $L_\mathrm{maj}$, and $T_\mathrm{b}$ were calculated 
using the ASURV survival analysis package \citep{ASURV}.
\begin{figure}
 \resizebox{\hsize}{!}{\includegraphics[clip=true,angle=-90]{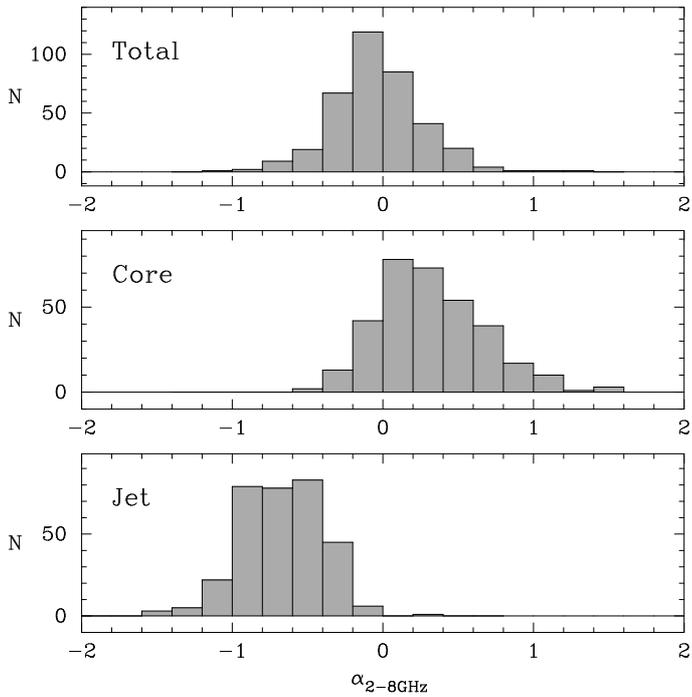}}
 \caption{
          Distributions of the 2.3--8.6~GHz spectral index of total VLBI scale emission ({\it top}), 
	  VLBI cores ({\it middle}), and jets ({\it bottom}).
         }
 \label{f:sp_index}
\end{figure}

\begin{table*}
\caption{VLBI core component properties.}
\label{t:Tb_core}
\begin{center}
\renewcommand{\footnoterule}{}
\begin{tabular}{cc|ccrr|ccrr}
\hline\hline
Source &  Epoch  &  \multicolumn{4}{c|}{2.3~GHz} & \multicolumn{4}{c}{8.6~GHz} \\
\cline{3-10}
    &     & $S_{\rm core}$ & $\theta_{\rm maj}$ & $\theta_{\rm min}$ & $T_{\rm b}$~~~~~ & $S_{\rm core}$ & $\theta_{\rm maj}$ & $\theta_{\rm min}$ & $T_{\rm b}$~~~~~ \\
    &     &      Jy        &     mas            &      mas           &     K~~~~~       &       Jy       &       mas          &      mas           &     K~~~~~       \\
(1) & (2) &      (3)       &     (4)            &      (5)           &     (6)~~~~      &      (7)       &       (8)          &      (9)           &     (10)~~~      \\
\hline
J0006$-$0623 & 2002/03/06 & 1.53 & 1.87 &    1.74 &    1.48e+11 & 1.55 & 0.81 &    0.27 &    1.56e+11 \\
J0011$-$2612 & 1999/03/08 & 0.40 & 0.57 &    0.57 &    6.10e+11 & 0.33 & 0.35 & $<$0.35 & $>$8.92e+10 \\
J0017$+$8135 & 2002/03/06 & 0.56 & 1.00 &    0.60 &    9.60e+11 & 0.54 & 0.41 &    0.18 &    5.16e+11 \\
J0019$+$7327 & 1999/03/08 & 0.68 & 1.51 &    0.84 &    3.51e+11 & 0.14 & 0.52 &    0.52 &    2.36e+10 \\
J0022$+$0608 & 2002/01/16 & 0.44 & 1.19 &    0.62 & $>$1.39e+11 & 0.41 & 0.16 &    0.16 & $>$2.55e+11 \\
J0027$+$5958 & 2001/10/29 & 0.08 & 0.82 & $<$0.82 & $>$2.62e+10 & 0.08 & 0.20 & $<$0.20 & $>$2.97e+10 \\
J0035$+$6130 & 2002/12/11 & 0.09 & 1.93 & $<$1.55 & $>$6.97e+09 & 0.09 & 0.45 &    0.45 & $>$7.29e+09 \\
J0050$-$0929 & 2002/03/06 & 0.56 & 1.66 &    0.92 & $>$8.61e+10 & 0.74 & 0.55 &    0.19 & $>$1.19e+11 \\
J0059$+$0006 & 1999/05/10 & 0.29 & 1.41 &    1.41 &    5.83e+10 & 0.10 & 0.55 & $<$0.55 & $>$9.39e+09 \\
J0102$+$5824 & 2002/01/16 & 0.57 & 0.96 &    0.72 &    3.17e+11 & 1.54 & 0.09 & $<$0.07 & $>$6.89e+12 \\
\hline
\end{tabular}
\end{center}
Columns are as follows:
(1) source name (J2000.0),
(2) date of observations (YYYY/MM/DD),
(3) fitted Gaussian core flux density,
(4) FWHM major axis of the fitted core Gaussian,
(5) FWHM minor axis of the fitted core Gaussian,
(6) core brightness temperature,
(7)--(10) the same as (3)--(6), respectively.
Table~\ref{t:Tb_core} is published in its entirety in the electronic version of the {\it Astronomy \& Astrophysics}.
A portion is shown here for guidance regarding its form and content.
\end{table*}

\begin{table*}
\caption{Statistics of distributions in Figs.~\ref{f:flux_hists}--\ref{f:Tb_hists}.}
\label{t:statistics}
\begin{center}
\renewcommand{\footnoterule}{}
\begin{tabular}{cccccccccccc}
\hline\hline
Frequency, [GHz] & \multicolumn{3}{c}{$S_\mathrm{VLBI}$, [Jy]} & \multicolumn{3}{c}{$S_\mathrm{unres}$, [Jy]} & 
                   \multicolumn{3}{c}{$S_\mathrm{unres}/S_\mathrm{VLBI}$} & \multicolumn{2}{c}{$S_\mathrm{core}/S_\mathrm{VLBI}$} \\
\cline{2-3}\cline{5-6}\cline{8-9}\cline{11-12}
                 &      Mean     & Median &&      Mean     & Median &&      Mean     & Median &&      Mean     & Median \\
\hline
2.3              & $1.02\pm0.08$ &  0.76  && $0.47\pm0.03$ &  0.34  && $0.51\pm0.02$ &  0.51  && $0.72\pm0.02$ &  0.75  \\
8.6              & $1.04\pm0.10$ &  0.69  && $0.54\pm0.05$ &  0.32  && $0.52\pm0.02$ &  0.51  && $0.73\pm0.02$ &  0.75  \\
\hline\hline\
                 &  \multicolumn{3}{c}{$S_\mathrm{core}$, [Jy]} & \multicolumn{3}{c}{$\theta_\mathrm{core}$, [mas]} & 
		    \multicolumn{3}{c}{$L_\mathrm{core\, proj}$, [pc]} & \multicolumn{2}{c}{$\log(T_\mathrm{b,\ core}$, [K])} \\
\cline{2-3}\cline{5-6}\cline{8-9}\cline{11-12}		 
		 &      Mean     & Median &&      Mean     & Median &&      Mean     & Median &&      Mean      & Median \\
\hline
2.3              & $0.69\pm0.05$ &  0.49  && $1.18\pm0.09$ &  1.04  && $7.56\pm0.66$ &  6.75  && $11.68\pm0.05$ &  11.40 \\
8.6              & $0.76\pm0.08$ &  0.47  && $0.33\pm0.03$ &  0.28  && $2.18\pm0.24$ &  1.90  && $11.69\pm0.07$ &  11.41 \\
\hline
\end{tabular}
\end{center}
\end{table*}

\subsection{Spectral properties}
\label{s:sp}

Flux density and parsec-scale structure variations in AGN jets require the observations
to be carried out simultaneously at different frequencies in order to properly investigate
the spectral properties of the sources. The RDV observations perfectly fulfill this requirement
and enable us to study spectral characteristics of the observed sources by using (i) a detailed 
reconstruction and analysis of a spectral index distribution map of a source and (ii) applying 
statistical methods to derive the general spectral properties and possible correlations/trends. 
In this paper, we implemented mostly the latter approach. For the statistical analysis, we used 
the calculated values of total VLBI flux densities from the 2.3~GHz and 8.6~GHz images (see 
Table~\ref{t:map_parameters}) to derive the integrated spectral index. The vast majority of 
sources have flat spectra, $\alpha>-0.5$, for more than 95\% of the sample (Fig.~\ref{f:sp_index}, 
top). 

We also derived median spectral indices for the VLBI jets by applying the following procedure.
First, for each source the X- and S-band images were obtained with the same pixel size (0.2~mas) 
and the same resolution by convolving them with the average restoring beam. A spectral index map 
was obtained by aligning the optically thin parts of the jet. Using the X-band data, the total 
intensity jet ridge line was then constructed with nearly equally spaced points separated by a 
distance of the pixel size. The procedure stops where the peak of a Gaussian curve fitted to the 
transverse jet profile becomes fainter than $4\mathrm{rms}$ noise level of the image. The position 
of the core component taken from the model fit was chosen as a starting point. We note that the 
VLBI core is often but not always the brightest feature on the map. Thus, the image phase center,
where the CLEAN algorithm tends to place the brightest component, would be an incorrect choice of 
starting point in such cases. Applying the ridge line to the spectral index distribution map, we 
extracted the values of $\alpha$ along the jet. After that, we masked out the core region within 
$r_\mathrm{core}<0.5(b_\phi^2+d_\phi^2)^{0.5}$, where $b_\phi$ and $d_\phi$ are the beam and 
fitted core sizes, respectively, along a position angle $\phi$ of the inner jet. Finally, the 
jet spectral index was estimated as the median of the remaning values. The described method provided 
results for 319 sources. The other 51 objects do not have enough structure, making the application 
of the procedure impossible. The measurements of the jet spectral indices are mostly distributed 
accross a range from $-0.2$ to $-1.2$ with the median value of $\alpha_\mathrm{2-8}^\mathrm{jet}=-0.68$, 
indicating optically thin radiation (Fig.~\ref{f:sp_index}, bottom). The spectral index of optically 
thin synchrotron radiation parametrizes the energy spectrum of relativistic radiative particles. 
Assuming a power-law energy distribution $N(E)=N_0E^{-\gamma}$, the power index $\gamma=1-2\alpha$ 
has the median value of $\gamma=2.4$ in the jets from the studied sample. The only positive value 
of a median jet spectral index was registered for the quasar J0927+3902, which has a short 
($\sim$3~mas) jet with a resolved feature several times brighter than the core at both frequencies. 
This feature that dominates in the structure is optically thick at 2.3~GHz and 8.6~GHz and has a 
spectral index of 0.36. We also refer to \cite{RATAN_spectra} for the simultaneous total flux 
spectrum of the object.

\begin{figure*}
\centering
 \resizebox{0.75\hsize}{!}{\includegraphics[clip=true,angle=-90]{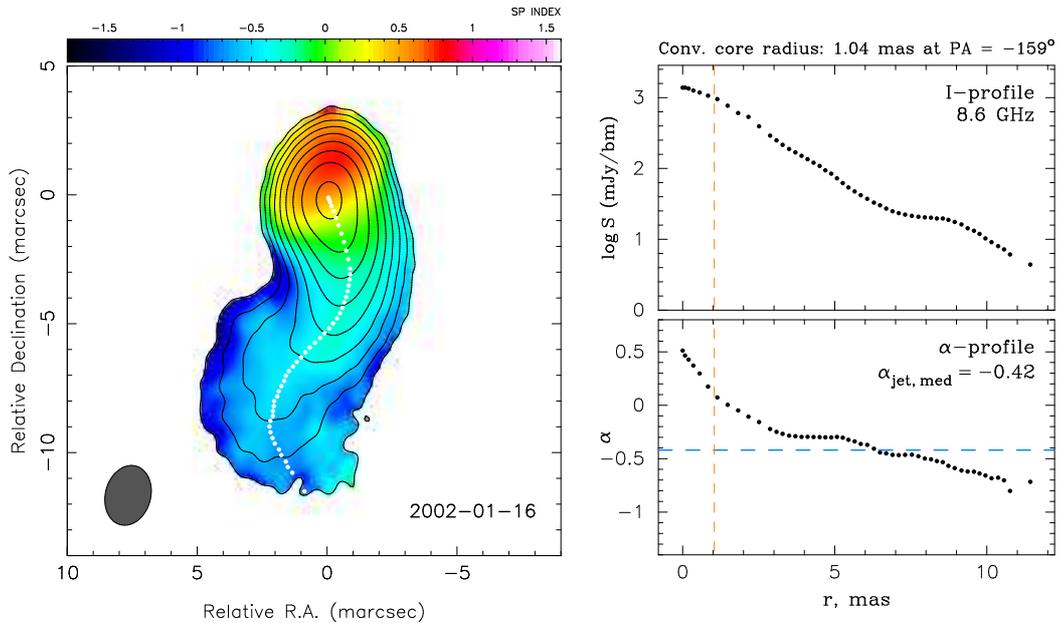}}
 \caption{Spectral index distribution in J2202+4216 (left) calculated between 2.3~GHz and 8.6~GHz with the
          8.6~GHz total intensity contours overlaid. White dots represent the total intensity ridge line, 
          along which we plot the profiles of total intensity (top right) and spectral index (bottom right).
          The lowest contour is plotted at 0.18\% of the peak brightness of 1395~mJy~beam$^{-1}$. The shaded
          ellipse represents the FWHM of the restoring beam of $2.34\times1.74$~mas at ${\rm PA}= -14.0\degr$,
          derived as the average between the corresponding interferometric restoring beams at 2.3~GHz and 
          8.6~GHz. The vertical orange dashed line shows the convolved radius of the VLBI core calculated 
          along the inner jet direction.}
 \label{f:sp_aging_BLLAC}
\end{figure*}

\begin{figure}
 \resizebox{0.9\hsize}{!}{\includegraphics[clip=true,angle=-90]{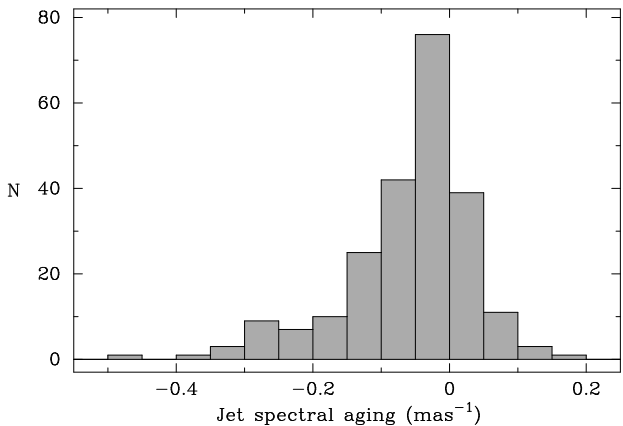}}
 \caption{Distribution of spectral index gradient along the total intensity ridge line for 228 sources.}
 \label{f:sp_aging_hist}
\end{figure}

The spectral index on average decreases along the jet ridge line (Fig.~\ref{f:sp_aging_BLLAC}) owing 
to the spectral aging caused by radiative losses of electrons. It slightly flattens in bright jet 
knots, which might be explained by shock acceleration at least in some cases. Using a linear least 
squares method, we estimated the spectral index gradients in the jets of 228 sources 
(Fig.~\ref{f:sp_aging_hist}). This analysis was done only for sources with at least 10 ridge line 
points beyond the core region. The mean value of the spectral index gradient is 
$-0.06^{+0.07}_{-0.08}\ \mathrm{mas}^{-1}$, where the errors are given at the 68\% confidence level. 

Spectral indices for the VLBI cores were obtained as a value of the $\alpha$-map at pixels corresponding 
to the starting point of the ridge line (X-band core). The distribution of the core spectral indices 
(Fig.~\ref{f:sp_index}, middle) has the median value of $\alpha_{2-8}^\mathrm{core}=0.28$, reflecting 
that the cores are partially opaque.

\begin{figure}
 \resizebox{0.9\hsize}{!}{\includegraphics[clip=true,angle=-90]{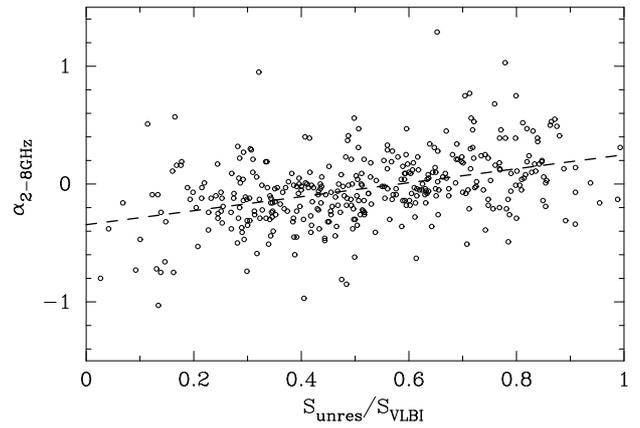}}
 \caption{
          Integral spectral index versus milliarcsecond compactness index at 8.6~GHz. 
	  The trend reflects the tendency for steep spectrum sources to be less compact.
	 }
 \label{f:sp_vs_comp}
\end{figure}

We also examined the connection between total spectral index and milliarcsecond compactness. The 
results for X-band data are shown in Fig.~\ref{f:sp_vs_comp}, the S-band dependence is qualitatively 
similar. The non-parametric Kendall's $\tau$-test confirms a positive correlation at a confidence 
level 99.9\%. The explanation of this visible trend is that sources with steeper spectra are expected
to have lower compactness indices, because a significant fraction of their VLBI flux should
originate from optically thin extended jet emission.

\begin{figure}
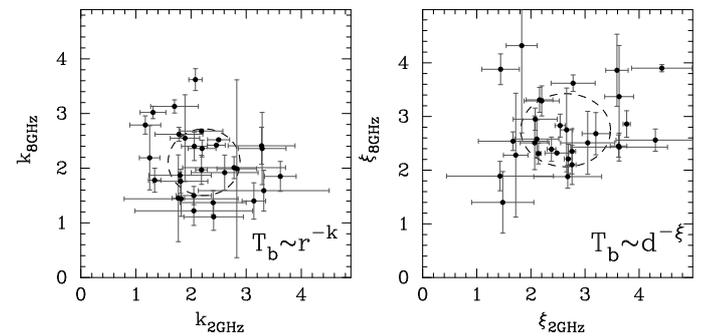

 \centering
 \includegraphics[width=0.24\textwidth,clip=true]{FIGS/fig15_k_index.eps}
 \includegraphics[width=0.24\textwidth,clip=true]{FIGS/fig15_xi_index.eps}
 \caption{
          8.6 versus 2.3~GHz power-law indices of brightness temperature gradients with distance to
          the core ({\it left}) and with jet component size ({\it right}). 
          dashed ellipse represents the $1\sigma$ error areas of the $k$- and $\xi$-index distributions.
         }
 \label{f:ks_indices}
\end{figure}

\begin{figure}[t!]
 \resizebox{\hsize}{!}{\includegraphics[clip=true,angle=-90]{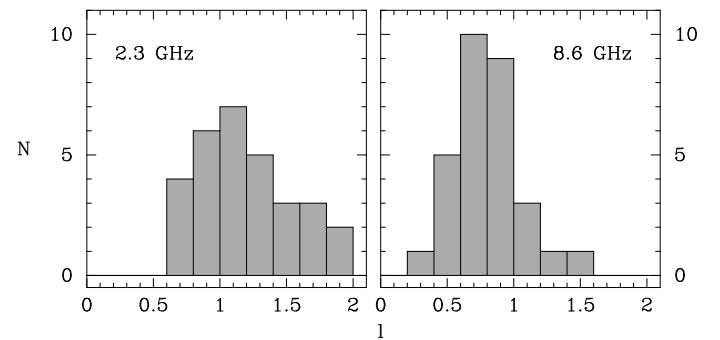}}
 \caption{
          Distributions of a power index $l$ assuming a $d\propto r^l$ dependence, where $d$ is the 
          transverse width of the jet at a distance $r$ from its apparent origin at 2.3~GHz ({\it left}) 
          and 8.6~GHz ({\it right}).
         }
 \label{f:jet_jeometry}%
\end{figure}

\subsection{Brightness temperature evolution along jets}
\label{s:Tb}
The sources with prominent jets were of particular interest to us, since they allow us to investigate 
the evolution of brightness temperature along the jet both as a function of distance to the core $r$, 
and the component's size $d$. Typically, a blob of relativistic plasma, which is detected on a VLBI 
map as a bright jet component, loses a substantial amount of energy through synchrotron radiation and 
adiabatic expansion and quickly becomes dim propagating down the outflow and increasing in size. 
These factors lead to the rapid decrease in brightness temperature along the jet. The evolution of 
$T_\mathrm{b}$ can be well-fitted by a simple power-law functions $T_\mathrm{b}\propto r^{-k}$ and 
$T_\mathrm{b}\propto d^{-\xi}$. We analyzed the brightness temperature gradients for 30 sources 
with a rich jet structure consisting of at least three model fitted jet components at both frequencies. 
The power-law index $k$ varies between 1.2 and 3.6, with the average value of 
$k_\mathrm{8\,GHz}\approx k_\mathrm{2\,GHz}=2.2\pm0.1$. The power-law index $\xi$ varies between 1.4 
and 4.4, with the average value of $\xi_\mathrm{8\,GHz}=2.7\pm0.1$ and $\xi_\mathrm{2\,GHz}=2.6\pm0.1$. 
Distributions of the fitted power indices $k$ and $\xi$ are shown in Fig.~\ref{f:ks_indices}. The 
power indices $k$ and $\xi$ can also be used to test jet models, e.g. adiabatic expansion in a 
shock-in-jet model suggested by \cite{Marscher90}.

Assuming that the width of a jet has a power-law dependence on distance as $d\propto r^l$, the 
power-law index $l$ can be readily estimated as $k/\xi$. The obtained distributions of $l$ 
(Fig.~\ref{f:jet_jeometry}) have median values of 1.2 at S-band and 0.8 at X-band, suggesting 
that jet regions probed by the RDV observations at 8.6~GHz are still collimating and therefore 
accelerating, while at 2.3~GHz the jets are more freely expanding and switch to a deceleration 
regime. This is consistent with the positive parallel accelerations detected by \cite{RDV_10yr} 
analyzing ten-year jet kinematics based on the 8.6~GHz RDV data and also \cite{Homan_accel} for 
15~GHz MOJAVE data. 

\subsection{Gamma-ray bright AGNs on parsec scales}

Among the 370 objects analyzed in this paper, 147 sources (40\%, see Table~\ref{t:general_info})
have been positionally associated with $\gamma$-ray detections within their corresponding 95\%
confidence region  made by the Large Area Telescope (LAT) on board the {\it Fermi} Observatory at 
energies above 100~MeV within two years of scientific operations \citep{2FGL}. The list of the 147
sources include 99 of 251 quasars (40\%), 39 of 46 BL Lacs (85\%), 6 of 31 radio galaxies (20\%), 
and 3 of 42 optically unidentified sources (7\%). A significantly higher detection rate for BL~Lacs 
is attributed to their harder spectra, which allows them to be detected more easily by the {\it Fermi} 
LAT \citep{2FGL_AGN}. We note that the vast majority of $\gamma$-ray detections associated with AGNs 
have significant radio flux densities on parsec scales as indicated by the highly successful rate of 
identifications of the {\it Fermi} LAT objects with a radio parsec-scale catalog of extragalactic jets 
\citep{Kovalev_FERMI_VLBI_assoc}. The {\it Fermi} era has already heralded a number of important links 
between $\gamma$-ray emission and parsec-scale properties of AGNs. It has been shown that LAT-detected 
sources are brighter and more luminous \citep{MF2}, have higher apparent jet speeds \citep{MF1,RDV_10yr}, 
higher Doppler factors \citep{MF4}, and characterized by wider apparent opening angles \citep{MF3}. 
There is also substantial evidence that $\gamma$-ray emission likely originates at the base of 
relativistic jet close to parsec-scale radio core \citep[e.g.,][]{MF2,Jorstad10_gamma,MF5}.

\begin{figure}
 \resizebox{\hsize}{!}{\includegraphics[clip=true,angle=-90]{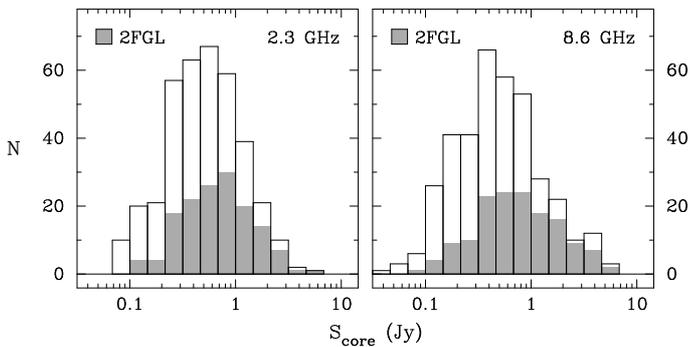}}
 \caption{
          Flux density distributions of VLBI core components at 2.3~GHz ({\it left}) and 8.6~GHz
          ({\it right}). The shaded areas represent LAT-detected (2FGL) sources.
         }
 \label{f:core_flux_Fermi_RDV}%
\end{figure}

\begin{figure}
 \resizebox{\hsize}{!}{\includegraphics[clip=true,angle=-90]{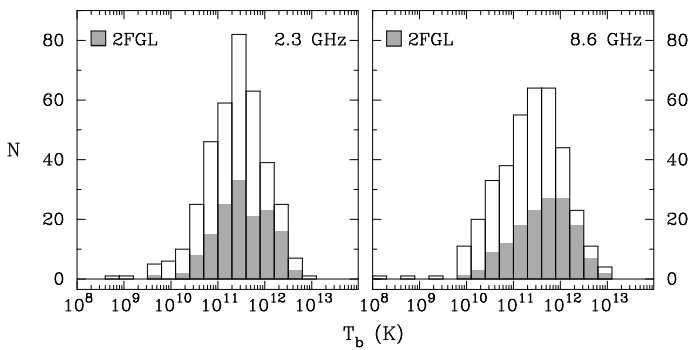}}
 \caption{
          Brightness temperature distributions of VLBI core components at 2.3~GHz ({\it left}) and
          8.6~GHz ({\it right}). The shaded areas represent LAT-detected (2FGL) sources.
         }
 \label{f:tb_core_Fermi_RDV}%
\end{figure}

We compared the flux densities of the VLBI core components for the LAT detected versus non-LAT detected RDV
sources (Fig.~\ref{f:core_flux_Fermi_RDV}) and found that the $\gamma$-ray bright sources have higher radio
core fluxes, with a mean value of 0.89~Jy versus 0.56~Jy at 2.3~GHz and 1.08~Jy versus 0.55~Jy at 8.6~GHz. 
In both cases, the Kolmogorov-Smirnov test indicates a probability $p<0.001$ that the corresponding samples 
are drawn from the same parent population. To complete a more detailed investigation of the $\gamma$-ray/radio 
jet relation, we also examined the core brightness temperature $T_\mathrm{b,\ core}$. Gehan's generalized 
Wilcoxon test from the ASURV survival analysis package indicates, at a $>99.9$\% confidence, that the 
$T_\mathrm{b,\ core}$ values for LAT-detected sources are statistically higher than those for the rest of 
the sample at 8.6~GHz (Fig.~\ref{f:tb_core_Fermi_RDV}, right), which is indicative of higher Doppler-boosting 
factors. For the 2.3~GHz data, the difference is marginal (92.3\% confidence) possibly due to the different 
spectral properties of the radio core and innermost jet components not being resolved with lower frequency 
observations. These findings confirm the early results reported by \cite{MF2}.

\begin{figure}[h]
 \centering
 \resizebox{0.95\hsize}{!}{\includegraphics[clip=true,angle=-90]{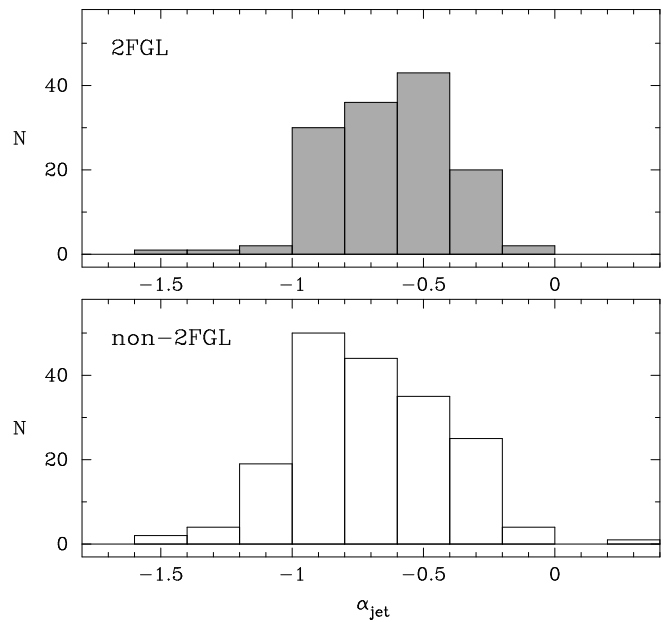}}
 \caption{
          Distribution of the 2.3--8.6~GHz spectral index for the VLBI jets detected ({\it top}) and
          not detected ({\it bottom}) by the {\it Fermi} LAT. See discussion of the positive value 
          found for the quasar J0927+3902 in Sec.~\ref{s:sp}.
         }
 \label{f:alpha_Fermi_RDV}%
\end{figure}

Additionally, we examined the distribution of the median spectral indices for the VLBI jets obtained
as described in \S~\ref{s:sp}. The distributions of $\alpha_\mathrm{jet}$ for 135 LAT-detected and 184 
non-LAT-detected sources (Fig.~\ref{f:alpha_Fermi_RDV}) differ at the significance level of 99.8\% as 
indicated by a K-S test. The corresponding medians are 
$\alpha_\mathrm{LAT\_Y}=-0.60$ and $\alpha_\mathrm{LAT\_N}=-0.72$, which correspond to the power indices
$\gamma_\mathrm{LAT\_Y}=2.20$ and $\gamma_\mathrm{LAT\_N}=2.44$ of the energy spectrum of relativistic 
radiative particles. This difference can be explained by a larger fraction of high-energy 
synchrotron-emitting electrons in $\gamma$-ray bright AGNs being produced by either more energetic or 
more frequent flares.

\section{Conclusions}
\label{s:summary}

We have obtained single-epoch images for a sample of 370 AGNs based on the global VLBI observations 
carried out simultaneously at 2.3~GHz and 8.6~GHz with the participation of ten VLBA antennas and 
up to ten additional geodetic radio telescopes. The parameters of the Gaussian model components 
representing the source structure have also been determined. The sample contains 251 quasars, 46 
BL~Lacertae objects, 31 radio galaxies, and 42 optically unidentified sources. At least 97\% of 
the sources manifest one-sided core-jet structures suggesting small angles to the line of sight. 
Even on milliarcsecond scales, the sources are highly compact and strongly core-dominated. Almost 
one-fourth of the core components are completely unresolved on the longest baselines of the global 
VLBI observations. 

Of the 370 observed sources, 96\% have flat-spectrum partially opaque cores with a median value of 
spectral index of $\alpha_{\mathrm{core}}\sim0.3$. The jet components are usually optically thin with 
a median value of $\alpha_{\mathrm{jet}}\sim-0.7$, which corresponds to $\gamma\sim2.4$ assuming a
power-law energy distribution $N(E)=N_0E^{-\gamma}$ of radiative particles. The spectral index on 
average decreases along the jet ridge line due to the spectral aging. It slightly flattens in bright 
jet knots, which might be explained by shock acceleration at least in some cases. We have found that 
the estimated brightness temperatures of the VLBI cores are affected by Doppler boosting, their typical 
values are found at the level of $10^{11}$~K, reaching up to $10^{13}$~K in rare cases. We note that 
40\% of measurements yield only lower limits either because of an unknown redshift or unresolved core 
component. No systematic difference was registered for values of $T_\mathrm{b,\ core}$ obtained at 
2.3~GHz and 8.6~GHz. 
 
The evolution of the brightness temperature of the jet components as a function of distance to the core, 
$r$, and transverse size, $d$, was fitted with power-law dependencies of $T_\mathrm{b}\propto r^{-k}$ 
and $T_\mathrm{b}\propto d^{-\xi}$, respectively, for 30 sources with straight and prominent jets, for 
which the mean values of $k_\mathrm{8\,GHz}\approx k_\mathrm{2\,GHz}=2.2$, $\xi_\mathrm{8\,GHz}=2.7$, 
and $\xi_\mathrm{2\,GHz}=2.6$ can be used to test jet models. On the scales probed by the 8.6~GHz 
observations, a jet geometry for these sources shows evidence of further jet collimation, while on the 
larger scales probed by the 2.3~GHz observations, the outflows expand more freely, perhaps switching 
into the deceleration regime.

Among the 370 presented sources, 147 (40\%) that are positionally associated with high-confidence 
$\gamma$-ray detections from the {\it Fermi} LAT Second Source Catalog have been found to have higher 
core flux densities and brightness temperatures suggesting preferentially higher Doppler-boosting 
factors. Furthermore, we have found that the LAT-detected AGNs are characterized by a less steep 
radio spectrum of the optically thin jet emission, most probably due to a larger fraction of 
high-energy radio emitting electrons being produced by flares.

Our acquired VLBI images together with their respective Gaussian models may be used for future 
astrometric projects to improve the accuracy of the source positions by taking into account their 
milliarcsecond structure. The highly compact bright sources would be potentially the most suitable 
targets for space VLBI studies with the {\it RadioAstron} mission \citep{RadioAstron}.

\begin{acknowledgements}

We would like to thank R.~Porcas, A.~Lobanov, L.~Petrov, T.~Savolainen, and G.~Piner 
for useful discussions. This work is based on the analysis of global VLBI observations including the VLBA, 
the raw data for which were provided to us by the NRAO archive. The National Radio Astronomy Observatory 
is a facility of the National Science Foundation operated under cooperative agreement by Associated 
Universities, Inc. YYK is partly supported by the Russian Foundation for Basic Research 
(project 11-02-00368), Dynasty Foundation, and the basic research program ``Active processes in galactic 
and extragalactic objects'' of the Physical Sciences Division of the Russian Academy of Sciences. This 
research has made use of NASA's Astrophysics Data System. This research has made use of the NASA/IPAC 
Extragalactic Database (NED), which is operated by the Jet Propulsion Laboratory, California Institute 
of Technology, under contract with National Aeronautics and Space Administration.

\end{acknowledgements}

\bibliographystyle{aa}
\bibliography{pab}

\end{document}